\documentclass[floatfix,twocolumn,aps,prl,amsmath,amssymb]{revtex4}
\usepackage[latin1]{inputenc}
\usepackage{graphicx}
\usepackage{dcolumn}
\usepackage{bm}
\usepackage{color}
\usepackage{url}
\usepackage[colorlinks=true ,citecolor=blue]{hyperref}

\begin{document}

\title{Superconducting and Pseudogap Transition Temperatures in High-Tc Cuprates and the $T_{c}$ Dependence on Pressure }
\author{E. C. Marino$^1$, Reginaldo O. Corr\^ea Jr$^{1}$, R. Arouca$^1$, Lizardo H. C. M. Nunes$^{2}$, Van S\'ergio Alves$^{3}$ } 
\affiliation{$^1$Instituto de F\'\i sica, Universidade Federal do Rio de Janeiro, C.P. 68528, Rio de Janeiro RJ, 21941-972, Brazil \\$^2$Departamento de Ci\^encias Naturais, Universidade Federal de S\~ao Jo\~ao del Rei, 36301-000 S\~ao Jo\~ao del Rei, MG, Brazil \\
$^3$Faculdade de F\'\i sica, Universidade Federal do Par\'a, Bel\'em PA, 66075-110, Brazil. }

\date{\today}

\begin{abstract}
We derive universal analytic expressions for the critical temperatures of the superconducting (SC) and pseudogap (PG)  transitions of the high-Tc cuprates as a function of doping. These are in excellent agreement with the experimental data both for single-layered materials such as LSCO, Bi2201 and Hg1201 and multi-layered ones, such as Bi2212, Bi2223, Hg1212 and Hg1223. Optimal doping occurs when the chemical potential vanishes. We show that the SC coupling is enhanced with the number of layers, N, which allows for an accurate description of Tc in the $Bi$, $Hg$ and $Tl$ multi-layered families of cuprates. We also study the pressure dependence of the SC transition temperatures, obtaining excellent agreement with the experimental data for different materials and dopings.
These results are obtained from an effective Hamiltonian for the itinerant oxygen holes, which includes both the electric repulsion between them and their magnetic interactions with the localized copper ions. We show that the former interaction is responsible for the PG and the latter, for the SC  phases, the phase diagram of cuprates resulting from the competition between both. The Hamiltonian is defined on a bipartite oxygen lattice, which results from the fact that only the $p_x$ and $p_y$ oxygen orbitals alternatively hybridize with the $3d$ copper orbitals.
From this, we can provide an unified explanation for the $d_{x^2-y^2}$ symmetry of both the SC and PG order parameters and obtain the Fermi pockets observed in ARPES experiments.

\end{abstract}

\maketitle

{\bf 1) Introduction}

Understanding the mechanism of high-Tc superconductivity in the cuprate materials is, at the same time, one of the most fascinating and challenging problems in physics. Thirty years after the experimental discovery of superconductivity in such materials \cite{bm}, we still have to face several fundamental phenomenological issues of the high-Tc cuprates, which cannot be properly accounted for by an underlying theory, despite the enormous amount of experimental, theoretical and numerical attempts made in that direction \cite{htsc0,htsc1,htsc2,htsc3,htsc4,htsc5}.
 
To mention just a few of these issues, let us recall that so far, the specific analytic expression for the curves representing the SC transition temperature as a function of doping, namely, $T_c(x)$, which form the characteristic SC domes in all high-Tc materials, is not known. Also, a theoretical framework that could provide an accurate analytical expression for the pseudogap transition temperature $T^*(x)$ is also not available. Furthermore, the detailed theoretical understanding of how the pressure influences and modifies the phase diagrams of high-Tc cuprates is still missing.

Concerning multi-layered cuprates, we still do not have an explanation
 for the fact that the optimal transition critical temperature increases as a function of the number of adjacent $CuO_2$ planes, up to a point and then stabilizes, as one can observe, for instance, in the, $Bi$, $Hg$ and $Tl$ families of cuprates
\cite{honma,honma0,honma1}. 
 
In this study, we address all the above phenomenological issues of high-Tc cuprates and provide explanations for each of them, which are in excellent agreement with the experimental data.
From the very outset it becomes clear that the BCS paradigm does not apply to the superconductivity found in cuprate materials. According to this paradigm, for two electrons to form a Cooper pair, their energies must differ by an amount less than the Debye energy in order for their mutual interaction, mediated by phonons, to become attractive ($|\epsilon_1 - \epsilon_2|< \hbar \omega_D$). This condition is met in a metal in the situation when most of the electrons are close to the Fermi surface ($|\epsilon - \epsilon_F|< \hbar \omega_D$). This situation, however, can only occur at very low temperatures, a fact that explains why the SC transition temperatures are so low in BCS superconductors. In the cuprates, the parent compounds are actually insulators and this fact itself points towards an alternative mechanism, which does not require such low temperatures.

The observation of the pseudogap phase which is marked by a suppression of the spectral weight  and the absence of a Fermi Liquid state, except at the very high doping regime, shows, conversely, that also the normal state of the high-Tc cuprate materials, above $T_c$, is far more complex than that in a conventional superconductor. Finally, the absences of the isotope effect and of the softening of phonon modes have strongly indicated that these materials cannot be described by the regular BCS-Theory, at least in what concerns the mechanism that produces an attractive electron-electron interaction.

 The existence of a strongly ordered antiferromagnetic phase in the parent compounds, which in the case of multi-layered cuprates may even coexist with the SC phase in descendent materials, has suggested, from the early days of high-Tc superconductivity, that an interplay between the magnetic interactions of the system and the mechanism of Cooper pair formation should be at the roots of superconductivity in cuprates.

The parent compounds are magnetically ordered Charge-Transfer Insulators, whose insulating nature and magnetic order are destroyed upon doping, thus suggesting that the magnetic-order-destroying Metal-Insulator transition that takes place in these systems before the onset of the SC phase may have some influence in the SC mechanism. 

The Hubbard model \cite{fazekas} is a paradigm in strongly correlated electronic systems, especially in what concerns the mapping of such systems onto magnetic models. Given the important role that magnetic interactions are believed to play in high-Tc superconductivity, many theoretical approachs to the high-Tc SC in cuprates are based on variations or approximations of the Hubbard model. Among these, the $t-J$ model \cite{zhangrice}, in the context of Anderson's Resonance Valence Bond (RVB) theory \cite{RVB1, RVB2}, has been extensively used in the literature, as can be seen in \cite{RVB3}.
One should also mention the so-called Spin Fluctuation Model, proposed by Monthoux, Balatsky and Pines, in which the SC pairing is mediated by the so-called paramagnons \cite{mbp}.

It seems clear, anyway, that the main physics in the high-Tc cuprates occurs in the $CuO_2$ planes, involving the electrons in the $3d_{x^2-y^2}$ orbitals of copper ions and $2p_x$ and $2p_y$ orbitals of oxygen ions. In this context,
the Coulomb interactions among these are agreedly well captured by the so-called Three Bands Hubbard Model (3BHM) \cite{3bhm,3bhm1}, which describes, besides the different hopping possibilities, the $dd$, $pp$ and $pd$ Coulomb repulsive interactions among the corresponding electrons. 
 This model was intensively studied, mainly numerically \cite{num_3band_1, num_3band_2,num_3band_3} and showed that indeed the holes go into the oxygen orbitals, thus making the doped holes to represent the itinerant degrees of freedom of the $CuO$ planes.

Given the complexity of the 3BHM, however, simplified versions thereof were considered. Perhaps the simplest of these versions is the $t-J$ model \cite{zhangrice}. This, however, has been criticised because, being a one band model, it does not consider the hopping between oxygen orbitals. Altough it is argued that the energy separation between states connected by this hopping is very big, it is, nevertheless, smaller than the one between the singlet and triplet bound state \cite{clust}, as pointed out in \cite{thesis_3band}. 

 A particularly interesting simplified version of the 3BHM is the Spin-Fermion Model (SFM) \cite{sf} which is formulated in terms of the holes doped into the $2p_x$ and $2p_y$ orbitals of oxygen ions and the electrons in the $3d_{x^2-y^2}$ orbitals of copper ions, which are localized as a consequence of the strong Coulomb repulsion, thus forming a Charge Transfer Insulator \cite{cti1,cti2,cti3,3bhm}.

The SFM describes, besides the hopping of holes, their Kondo-like magnetic interaction with the localized spins of the copper ions,with coupling $J_K$, as well as the antiferromagnetic (AF) super-exchange interaction between the latter, with coupling $J_{AF}$, which produces a N\' eel ground state on the parent compounds, in the undoped limit. The SFM does not include the description of the Coulomb repulsion between the doped holes, which is quite strong. Hence a more complete model, which descends from the 3BHM would contain, besides the magnetic interactions of the SFM, the Coulomb repulsion among the doped holes, which is described by a pure Hubbard-like interaction. The resulting model may be called Spin-Fermion-Hubbard Model (SFHM) \cite{emsc}.

This model ,
defined on the oxygen lattice of the $CuO_2$ planes of the cuprates will be our starting point for deriving a Hamiltonian, which can provide an acceptable description of the SC in cuprates. Indeed
the Hamiltonian that we propose as effectively describing the dynamics of the doped holes in cuprates is derived from the SFHM by means o two well-known operations: a) tracing out the localized copper ion spins; b) performing a second order perturbative expansion in $t_p/U_p$, where $t_p$ is the hopping parameter and $U_p$ the Hubbard local repulsion parameter of the $p$-orbital holes.

The first operation will yield a hole-attractive term with a coupling parameter $g_S=\frac{J_K^2}{8 J_{ AF}}$, whereas the second will produce a hole-repulsive term with a coupling parameter  $g_P=\frac{2t_p^2}{U_p}$, coming, respectively, from the magnetic and Coulomb repulsion terms of the SFHM.

 A crucial feature of our model, however, is the observation that the oxygen lattice breaks down into two inequivalent sublattices, for which the $p_x$ or $p_y$ oxygen orbitals, respectively, overlap with the copper $3d$ orbitals. 
Cooper pairs are formed by combining holes belonging to the two different sublattices, which contain respectively, $p_x$ and $p_y$ orbitals. This naturally leads to a d-wave SC order parameter, which is favored by the attractive interaction sector, which derives from the Kondo magnetic interaction between doped holes and localized spins. The term describing the repulsion between holes, conversely, favors the onset of a non-vanishing d-wave PG order parameter, which results from exciton (electron-hole pair, each belonging to a different sublattice) condensation.
 Our model naturally provides a unified explanation for the d-wave character both of the SC and PG order parameters, the latter leading to the DDW (d-density wave) scenario \cite{ddw,ddw0,ddw1} proposed to explain the PG phenomena.

 The picture that emerges from our study is that the phase diagram of the cuprates results from the duality \cite{ecm1} between the formation of Cooper pair and (DDW) exciton condensates, both with a d-wave symmetry. The two effective interaction terms contained in our Hamiltonian can be derived from the Hubbard-Spin-Fermion model \cite{emsc}, which is the starting point for the present approach. 

The doping mechanism is explicitly taken into account by the introduction of a constraint relating the fermion number to a function of the stoichiometric doping parameter. Since the relation between the stoichiometric chemical potential and the actual amount of charge doped into the $CuO_2$ planes is unknown in general, we adjust the value of the parameter determining the stoichiometric chemical potential, in order to fit the experimental data. This is the only fitted parameter in our approach to the cuprates. From it we can derive the values of the coupling parameters $g_S$ and $g_P$.
We find a remarkable  agreement between the numerical values obtained for these two coupling parameters: a) by expressing them in terms of the original parameters of the SFHM; and b) by expressing them in terms of the adjusted  phenomenological chemical potential, chosen to fit the experimental data for the cuprates. This strongly indicates the correctness of our model.

Quantum dynamical effects are brought up by
functional integrating out the fermion degrees of freedom. This allows the obtainment of the grand-canonical potential $\Omega(\Delta_0,M_0,\mu_0)$ in terms of the superconducting order parameter, $\Delta_0$, the pseudogap order parameter $M_0$, the chemical potential $\mu_0$ and the temperature. This includes the full quantum fluctuations of the holes' degrees of freedom. Then, minimizing the effective potential, which corresponds, to $\Omega$, we are able to verify that the occurrence of nonzero $\Delta_0$ and $M_0$ are, in general, mutually exclusive, thereby indicating a competition between the PG and SC phases. The only exception would occur for the case when $g_S=g_P$.

 By taking the limits $\Delta_0 \rightarrow 0$ and  $M_0 \rightarrow 0$, respectively, we capture the threshold for the SC and PG transition and thereby arrive at analytic expressions for the critical SC and PG temperatures as a function of doping, namely $T_c(x)$ and $T^*(x)$. These reproduce the familiar SC domes, as well as the PG lines found in the cuprates and is in excellent agreement with the experimental data, both for single-layered materials such as LSCO, Bi2201 and Hg1201 and for multi-layered ones, such as Bi2212, Bi2223, Hg1212 and Hg1223.
Our results indicate that the optimal amount of stoichiometric doping, $x_0$, which leads to the maximal $T_c$ occurs when the chemical potential vanishes: $\mu_0(x_0)=0$,  $T_c(x) \leq T_c(x_0)$. We find that, similarly to the BCS result, the optimal temperature, apart from a natural scale (which is the Debye energy in the case of BCS superconductors) contains a function of the coupling parameter $g_S$, which is non-analytical at $g_S\rightarrow 0$ and tends to $1$ for $g_S\rightarrow \infty$. The first fact indicates that our approach is completely non-perturbative.

The increase of the optimal temperature as we increase the number of adjacent planes in the primitive unit cell of multi-layered cuprates, can be simply understood within our approach, as we show that the effective coupling parameter $g_S$ is enhanced by the number, $N$, of such planes: $g_S \rightarrow Ng_S$. 

We finally study the effects of an applied external pressure on the SC transition temperature $T_c(x)$ as well as on the PG transition temperature $T^*(x)$ and show that pressure would strongly affect the former, however would not produce any effects on the latter. We obtain analytical expressions for $T_c(x,P)$, both for fixed values of doping and for fixed values of the pressure, in the latter case, describing the SC dome for different values of the pressure, always in excellent agreement with the experimental data. 

The article is organized as follows. In Section 1, we introduce the subject; in Section 2, we derive the Hamiltonian of the model; in Section 3, we describe the doping process and the formation of the Fermi surface; in Section 4, we obtain the effective potential and the general expressions for $T_c(x)$ and  $T^*(x)$; in Section 5, we apply the results to describe the phase diagram of several cuprate materials; in Section 6, we study the effects of the number of layers on $T_c(x)$ and in Section 7 we describe the effects of pressure on $T_c(x)$. Concluding Remarks are presented on Section 8. Three Appendices are also included.\\
\bigskip

{\bf 2) The Effective Hamiltonian}\\
\bigskip

{\bf 2.1) The Oxygen Sublattices}\\
\bigskip

\begin{figure}
	[h]
	\centerline{
		%\figurename{TN}
		\includegraphics[scale=0.3]{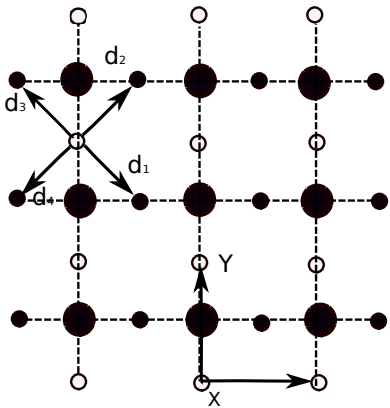}
		%clip, 
		%angle=90, 
		%width=1.\textwidth]
		%\include{TN}
	}
	\caption{The $CuO_2$ lattice. Big dark circles are $Cu$ ions. Small dark and white circles are $O$ ions. Notice that these form a bipartite lattice whose primitive vectors, $\textbf{X}$ and $\textbf{Y}$ are shown. These point along the x, y directions, which are aligned with the $Cu-Cu$ ions direction, according to our convention. We also show the $\textbf{d}_i$, $i=1,...,4$, vectors connecting the $O$ ions of a given sublattice to their counterparts in the complementary sublattice}
	\label{figa4}
\end{figure}

\begin{figure}
	[h]
	\centerline{
		%\figurename{TN}
		\includegraphics[scale=0.5]{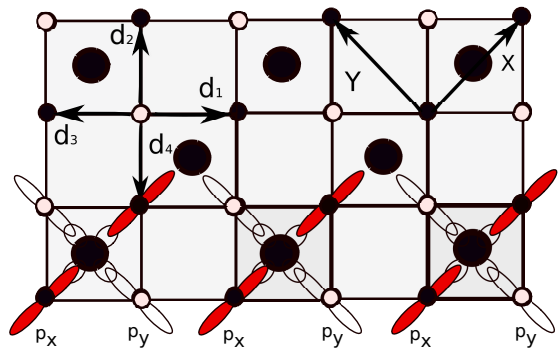}
		%clip, 
		%angle=90, 
		%width=1.\textwidth]
		%\include{TN}
	}
	\caption{The $CuO_2$ lattice: a 45 degree rotation view. We also represent, in the bottom, the $p_x$ and $p_y$ oxygen orbitals that alternatively hybridize with the copper 3d orbitals. The displayed vectors are described in Fig. \ref{figa4}.}
	\label{figb4}
\end{figure}

\begin{figure}
	[h]
	\centerline
	{
		%\figurename{TN}
		\includegraphics[scale=0.4]{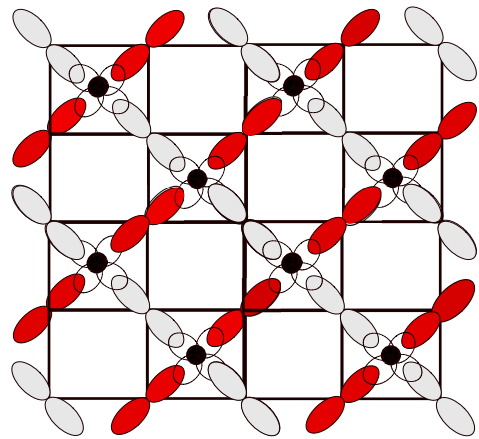}
		%clip, 
		%angle=90, 
		%width=1.\textwidth]
		%\include{TN}
	}
	\caption{The bipartite oxygen crystal structure, showing two sublattices of oxygen ions, formed, respectively, by $p_x$ (red) and $p_y$ (white) orbitals that overlap the $Cu^{++}$ d-orbital.  Black dots are the $Cu^{++}$ ions. The displayed vectors are described in Fig. \ref{figa4}.}
	\label{fig01}
\end{figure}

An outstanding feature of all High-Tc cuprates is the presence of one or more $CuO_2$ planes, intercepting the primitive unit cell of such compounds. The $CuO_2$ planes have a lattice structure in which $Cu^{++}$ ions occupy the sites and oxygen ions the links of a square lattice, with a lattice parameter $a=3.8 $\AA .
These ions are in a $3d^9$ electronic configuration, which results in one spin 1/2 per site. The system of copper ions is a Charge Transfer Insulator, hence, from this point of view, it forms an array of localized spins
interacting with the nearest-neighbors through the super-exchange mechanism. This structure is ultimately responsible for the antiferromagnetic properties observed in the high-Tc cuprates.  From the point of view of the oxygen ions, however, the picture is different. Indeed,
the oxygen ions are themselves, placed on the sites of a square lattice, with a lattice parameter $a'=\sqrt{2}  a / 2 = 1.9\sqrt{2}$ \AA \, which possesses two sublattices, containing, respectively, $p_x$ and $p_y$ oxygen orbitals, which overlap with the $Cu^{++}$ d-orbitals (see Figs. \ref{figa4}, \ref{figb4},\ref{fig01}), thereby forming bridges that will allow not only hole hopping along the whole oxygen lattice, but also the formation of Cooper pairs as well as excitons along these bridges. As we shall see, this fact naturally explains why both the SC and PG gaps have a d-wave symmetry.

An attentive analysis of the cuprates must consider the fact that the $Cu^{++}$ ions, which have four oxygen nearest neighbors, hybridize in different ways with the two of them placed along the $x$ and $y$ directions, thereby creating two inequivalent oxygen sublattices. Indeed, notice that
each oxygen ion possesses both one $p_x$ and one $p_y$ orbitals, however, only one of them alternately hybridizes with the copper 3d orbitals, hence forming oxygen sublattices, which have, respectively, either $p_x$ or  $p_y$ orbitals hybridized with the copper ions.
It follows that both the hopping and the interaction of the corresponding oxygen holes (see Figs. \ref{figa4}, \ref{figb4} \ref{fig01}), thereby assisted by the $Cu^{++}$ ions, must involve the two different $p_x$ and $p_y$ oxygen sublattices.

Our convention concerning the coordinate frame orientation is such that the $x$ and $y$ axes coincide with the $Cu-Cu$ ion directions and also with the primitive vectors of the two sublattices forming the bipartite oxygen lattice. In a square lattice, the reciprocal lattice primitive vectors are proportional to the original ones, hence our $k_x$ and $k_y$ directions are respectively parallel to the original $x$ and $y$ directions \cite{ecm2,rmp1}.

In Figs. \ref{figa4} and \ref{figb4}, $\textbf{d}_i$, $i=1,...,4$ are the vectors connecting every oxygen ion with its four nearest neighbors of the complementary sublattice and $\textbf{X}= a \hat{\textbf{x}},\textbf{Y}= a \hat{\textbf{y}}$ are the primitive vectors of the copper lattice and also of each of the oxygen sublattices, all of them with a lattice parameter $a$. Notice that the four vectors  $\textbf{d}_i$, $i=1,...,4$ are given by
\begin{eqnarray}
& &\textbf{d}_1=\frac{1}{2}[\textbf{X}-\textbf{Y}]\ ;\ \textbf{d}_2=\frac{1}{2}[\textbf{X}+\textbf{Y}]\nonumber \\
& &\textbf{d}_3=\frac{1}{2}[-\textbf{X}+\textbf{Y}]\ ;\ \textbf{d}_4=\frac{1}{2}[-\textbf{X}-\textbf{Y}] 
\label{11xa},
\end{eqnarray}

\begin{figure}
	[h]
	\centerline
	{
		%\figurename{TN}
		\includegraphics[scale=0.25]{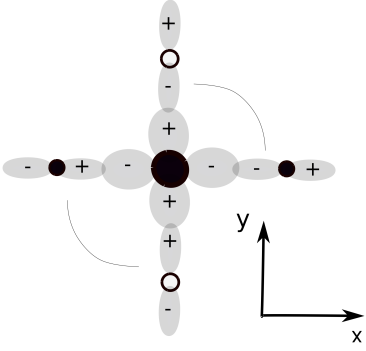}
		%clip, 
		%angle=90, 
		%width=1.\textwidth]
		%\include{TN}
	}
	\caption{The hybridization of $p$ and $d$ orbitals of the oxygen and copper ions. Notice the asymmetry between the x and y directions, revealed by the phase sign in the overlap of the $p$ and $d$ orbitals.}
	\label{figa5}
\end{figure}

In Fig. \ref{figa5} we represent the hybridization of the $p$ and $d$ orbitals in the $CuO_2$ planes. First notice that only one of the two oxygen $p$-orbitals contained in the plane hybridizes with the copper $d$-orbitals. This creates a bipartite oxygen lattice, a fact that has profound consequences on the superconducting and pseudogap order parameters, as we shall see. Notice the x-y antisymmetry under a 90 degrees rotation, produced by the signs of the overlapping orbitals. This has, as a consequence, the manifestation of a d-wave symmetry, both in the SC and PG order parameters.\\
\bigskip

{\bf 2.2) The Spin-Fermion-Hubbard Hamiltonian}\\
\bigskip

Our starting point is the following Hamiltonian that includes, besides the antiferromagnetic super-exchange interaction between neighbor localized copper ions, the magnetic interaction involving the localized and itinerant  magnetic dipole moments in the planes, as well as the local Coulomb repulsion among doped holes:
\begin{eqnarray}
H_{SFH}=H_{0}+H_{U}+H_{AF}+H_{K}
\label{00a},
\end{eqnarray} 
where  $H_{0}$ is the kinetic hopping term for the itinerant holes in oxygen ions, describing hops between different sublattices $A$ and $B$. $H_{AF}$ describes the antiferromagnetic interaction among the localized copper magnetic moments, whereas
$H_{K}$ is a Kondo-like magnetic interaction between the itinerant magnetic dipole moments of oxygen, corresponding to holes in sublattices $A$ and $B$ and the localized magnetic moments of copper. $H_U$ is the local Coulomb repulsion between doped holes.

In order to express the different Hamiltonian terms described above, we now introduce the hole creation operators for each of the two sublattices, namely, $ \psi^\dagger_{A\sigma}(\textbf{R})$ and
 $\psi^\dagger_{B\sigma}(\textbf{R}+\textbf{d}_i)$ where $\textbf{R}$ are the sites of the $A$ sublattice and $\textbf{d}_i, i=1...4$ the vectors connecting each site of the $A$ sublattice to the four nearest neighbors of $\textbf{R}$, belonging to the $B$ sublattice. $\sigma=\uparrow,\downarrow$ represent the holes' two spin orientations. Let us also represent by $\textbf{S}_I$ the localized spin operator of the copper ion placed on site $I$ of the square lattice formed by the copper ions in the $CuO_2$ planes. Each copper localized spin has four nearest neighbor hole sites, two of them in sublattice $A$ and other two in sublattice $B$.  In terms of these operators, we can express the four terms of the Hamiltonian above as follows:

\begin{eqnarray}& &
\hspace{-5mm}H_0 =-t_{p} \sum_{\textbf{R},\textbf{d}_i}\sum_{\sigma}\psi_{A,\sigma}^\dagger(\textbf{R})\psi_{B,\sigma}(\textbf{R}+\textbf{d}_i) +hc
\nonumber \\
&\ &
\hspace{-5mm}H_U =
%\nonumber \\
%&\ &
U_p \sum_{\textbf{R}} n^A_\uparrow n^A_\downarrow + U_p \sum_{\textbf{R}+\textbf{d}} n^B_\uparrow n^B_\downarrow 
\nonumber \\
& &
\hspace{-5mm}H_{AF} = J_{AF}\sum_{\langle IJ \rangle} \textbf{S}_I\cdot  \textbf{S}_J
\nonumber \\
& &
\hspace{-5mm}
H_K =J_K\sum_{I}\sum_{\textbf{R},\textbf{R}+\textbf{d}\in I}
 \textbf{S}_I\cdot\left[\mathcal{S}_{A}(\textbf{R}) +\mathcal{S}_{B}(\textbf{R}+\textbf{d})\right ] 
\label{0a}
\end{eqnarray}

In the above expression,
\begin{eqnarray}
& & \mathcal{S}_{A}(\textbf{R}) =\frac{1}{2}  \psi_{A\alpha}^\dagger(\textbf{R})\vec{\sigma}_{\alpha\beta}\psi_{A\beta}(\textbf{R}) 
\nonumber \\
& & \mathcal{S}_{B}(\textbf{R}+\textbf{d}) = \frac{1}{2}  \psi_{B\alpha}^\dagger(\textbf{R}+\textbf{d})\vec{\sigma}_{\alpha\beta}\psi_{B\beta}(\textbf{R}+\textbf{d}) 
\label{0b}
\end{eqnarray}
 are the spin operators for the holes in sublattices $A,B$; $t_{p}$ is the nearest neighbor oxygen-lattice hopping parameter,  $J_{AF}$ is the AF coupling between nearest neighbors of the copper ion lattice and $J_K$ is the Kondo magnetic coupling between the itinerant oxygen holes and the localized copper ions. $U_p$ is the local Coulomb repulsion for the holes in oxygen orbitals. In terms of the original Three Bands Hubbard model parameters, we have \cite{3bhm,3bhm1}
\begin{eqnarray}
J_{AF} = \frac{4 t^4_{pd}}{(\Delta_E+U_{pd})^2}\left[\frac{1}{U_d}+\frac{2}{2\Delta_E +U_p}\right ]
\label{0bx}
\end{eqnarray}
and
\begin{eqnarray}
J_{K} =  t^2_{pd}\left[\frac{1}{\Delta_E}+\frac{1}{U_d-\Delta_E}\right ].
\label{1bx}
\end{eqnarray}
For LSCO, the 3BHM parameters are \cite{dp} :  $U_d=8.5 \ eV$, $U_p = 5.5 \ eV$, $U_{pd}=0.897\ eV$, $t_p=0.91 \ eV$, $t_{pd}= 1.48\ eV$, $\Delta_E=\epsilon_p-\epsilon_d=2.75\ eV$, which imply $J_K=1.17\ eV$ and $J_{AF}=0.43\ eV$.

Since $U_d > \Delta_E$, we see that the energy split between the two Hubbard bands is larger than the energy separation between the $d$ and $p$ orbitals, thus characterizing the undoped system as a Charge Transfer Insulator, with a gap          $\Delta_E$ \cite{cti1,cti2,cti3}.
\\
\bigskip

{\bf 2.3) The Effective Hamiltonian for the Itinerant Holes}\\
\bigskip

In order to obtain an effective Hamiltonian for the itinerant degrees of freedom, we are going to perform two distinct familiar operations on the Hamiltonian (\ref{0a}). We first trace out the localized degrees of freedom, represented by the copper spins $S_I$. For this, we follow the usual procedure (see Appendix A; also \cite{ecm2,ecm1} for instance) which employs spin coherent states, in order to express the partition function as a functional integral over a classic unit vector field $\textbf{N}$, which replaces the localized spin operator $\textbf{S}_I/2$ in $H_{AF}$ and $H_K$. Then, after separating the antiferromagnetic fluctuations from the ferromagnetic ones, we integrate over the latter.
The second operation consists in performing a second order $t_{p}/U_p$ perturbative expansion in $H_0+H_U$ for the energy eigenvalues. Subsequently, we determine an effective hamiltonian such that, in its presence, the first-order corrections to the energy eigenvalues coincide with the previous second order result for such eigenvalues.

For tracing out the localized spins we start from the partition function, given by
\begin{eqnarray}
Z= {\rm Tr}_{\textbf{S}_I}{\rm Tr}_{\psi} e^{- \beta H_{SFH}[\textbf{S}_I,\psi]}, 
\label{a}
\end{eqnarray}
and, after performing the trace over $\textbf{S}_I$ (actually over the ferromagnetic fluctuations of $\textbf{S}_I$; see Appendix A), we obtain
\begin{eqnarray}
Z= Z_{NL\sigma M} {\rm Tr}_{\psi} e^{- \beta \Big[ H_0[\psi]+ H_U[\psi]+ H_1[\psi] \Big]},
\label{b}
\end{eqnarray}
where $ Z_{NL\sigma M}$ is the partition function of the Nonlinear Sigma Model ($NL\sigma M$), which describes the magnetic properties of the $Cu^{++}$ localized magnetic dipole moments. In such a description, these become proportional to the $NL\sigma$ field $\textbf{n}$. The magnetic part of the phase diagram of High-Tc cuprates can be derived from the $NL\sigma M$ \cite{ecmmbsn}.

In the process of tracing out the localized spins, we generate a new term, namely $H_1[\psi]$, which is given by

\begin{eqnarray}
H_1[\psi] =-\frac{J^2_K}{8 J_{ AF}}\sum_{ij} \left [  \mathcal{S}_{A,i}+\mathcal{S}_{B,j}\right ]^2  
\label{22a}
\end{eqnarray}
where the sites $i$ and $j$ belong, respectively, to the $A$ and $B$ oxygen sublattices.

Inserting (\ref{0b}) into (\ref{22a}), we obtain, up  to a constant (see Appendix B),

\begin{eqnarray}
H_1[\psi] = -\frac{J_K^2}{8 J_{ AF}} \left [ \Sigma - \Xi_1
\right ]
\label{3a}
\end{eqnarray}
where
\begin{eqnarray}
& & \Sigma =\psi^\dagger_{A\uparrow}\psi^\dagger_{B\downarrow}\psi_{B\uparrow}\psi_{A\downarrow}+\psi^\dagger_{A\downarrow}\psi^\dagger_{B\uparrow}\psi_{B\downarrow}\psi_{A\uparrow}
\nonumber \\
 & & \Xi_1 = \psi^\dagger_{A\uparrow}\psi_{A\uparrow}\psi^\dagger_{B\downarrow}\psi_{B\downarrow}+\psi^\dagger_{A\downarrow}\psi_{A\downarrow}\psi^\dagger_{B\uparrow}\psi_{B\uparrow}=
 n^A_+n^B_-n^A_-n^B_+.
\nonumber \\
\label{4a}
\end{eqnarray}

We now turn to the
perturbative expansion in $ H_0$, which is performed in Appendix C. The result is the replacement of $H_U $ for
 an additional effective interaction term for the itinerant holes, given by,
\begin{eqnarray}
H_2[\psi]= - \frac{2t_{p}^2}{U_p} \left [ \Pi - \ \Xi_2 \right ]
\label{4a}
\end{eqnarray}
where
\begin{eqnarray}
\Pi=\psi^\dagger_{A\uparrow}\psi_{B\uparrow}\psi^\dagger_{B\downarrow} \psi_{A\downarrow}
+\psi^\dagger_{A\downarrow}\psi_{B\downarrow}\psi^\dagger_{B\uparrow}\psi_{A\uparrow}
\label{4az}
\end{eqnarray}
and
\begin{eqnarray}
\Xi_2 = \psi^\dagger_{A\uparrow}\psi_{A\uparrow}\psi^\dagger_{B\uparrow}\psi_{B\uparrow}+\psi^\dagger_{A\downarrow}\psi_{A\downarrow}\psi^\dagger_{B\downarrow}\psi_{B\downarrow}=
  n^A_+n^B_+n^A_-n^B_-
\nonumber \\
\label{4b}
\end{eqnarray}
The first order perturbative corrections to the energy eigenvalues in the presence of $H_2[\psi]$, coincide with the second order ones obtained from $H_0+H_U $.

The total effective Hamiltonian for the itinerant holes, therefore, will be

\begin{eqnarray}
& &\hspace{-3mm}H_{eff}[\psi]=H_0+H_{1}+H_{2} 
\nonumber \\
\nonumber \\
& &\hspace{-3mm}H_{eff}[\psi]=-t \sum_{\textbf{R},\textbf{d}_i} \psi_{A\sigma}^\dagger(\textbf{R})\psi_{B\sigma}(\textbf{R}+\textbf{d}_i)+hc
\nonumber \\
& &\hspace{-3mm}-g_S\sum_{\textbf{R},\textbf{d}_i} \Big [\psi_{A\uparrow}^\dagger(\textbf{R})\psi_{B\downarrow}^\dagger(\textbf{R}+\textbf{d}_i)
 + 
\psi^\dagger_{B\uparrow}(\textbf{R}+\textbf{d}_i)\psi_{A\downarrow}^\dagger(\textbf{R})\Big]
\nonumber \\
& &\hspace{-3mm}\times
\Big [\psi_{B\downarrow}(\textbf{R}+\textbf{d}_i)\psi_{A\uparrow}(\textbf{R})
 +  \psi_{A\downarrow}(\textbf{R}) \psi_{B\uparrow}(\textbf{R}+\textbf{d}_i)\Big]
\nonumber \\
& &\hspace{-3mm}-g_P
\sum_{\textbf{R},\textbf{d}_i} \Big [\psi_{A\uparrow}^\dagger(\textbf{R})\psi_{B\uparrow}(\textbf{R}+\textbf{d}_i)
 +\psi_{A\downarrow}^\dagger(\textbf{R}) \psi_{B\downarrow}(\textbf{R}+\textbf{d}_i)
\Big]\nonumber \\
& &\hspace{-3mm}\times
\Big [\psi_{B\uparrow}^\dagger(\textbf{R}+\textbf{d}_i)\psi_{A\uparrow}(\textbf{R})
+\psi_{B\downarrow}^\dagger(\textbf{R}+\textbf{d}_i) \psi_{A\downarrow}(\textbf{R})\Big]
\label{0}
\end{eqnarray}
In the above expression,  $g_S$, is the hole-attractive interaction coupling parameter and $g_P$, the hole-repulsive one.
According to (\ref{3a}) and (\ref{4a}), we have
\begin{eqnarray}
 g_S = \frac{ J^2_K }{8 J_{AF}}   \ \ \ \; \ \ \  g_P = \frac{2 t_p^2}{U_p}
\label{01}
\end{eqnarray}

Using the values of the magnetic coupling parameters valid for LSCO, provided in Sect. 2.2, we have: $J_{K} =1.17 \ eV$ and 
$J_{AF} = 0.43 \ eV$. From this, we obtain 
\begin{eqnarray}
\frac{J^2_{K}}{8  J_{AF}}= 0.39793\ eV
 \label{001}
\end{eqnarray}

This corresponds, with excellent accuracy, to the value obtained below from experimental data for the LSCO cuprate: 
\begin{eqnarray}
g_S=0.39406\ eV.
\label{012}
\end{eqnarray}

Also, using $t_p = 0.91\ eV$, $U_p=5.50\ eV $, \cite{dp} we get 
\begin{eqnarray}
& &  \frac{2 t_p^2}{U_P}= 0.30113\ eV
\label{01}
\end{eqnarray}
 This corresponds, also with excellent accuracy, to the value obtained below from experimental data for the LSCO cuprate: 
\begin{eqnarray}
g_P= 0.30547\ eV.
\label{0001}
\end{eqnarray}
These remarkable agreements are a strong indication that
 we are correctly modeling the high-Tc cuprates.
\\
\bigskip

{\bf 2.4) Hubbard-Stratonovitch Fields. The SC and PG Order Parameters}\\
\bigskip

The Hamiltonian above can be written, up to a constant, in trilinear form, in terms of the Hubbard-Stratonovitch fields 
$\Phi$ and $\chi$, namely
\begin{eqnarray}
& \hspace{-3mm}H_{eff}=&-t_{p} \sum_{\textbf{R},\textbf{d}_i}\sum_{\sigma}\psi_{A,\sigma}^\dagger(\textbf{R})\psi_{B,\sigma}(\textbf{R}+\textbf{d}_i) +hc
\nonumber \\
& &\hspace{-3mm}\begin{split}+ \sum_{\textbf{R},\textbf{d}_i}\Phi(\textbf{d}_i) 
\Big[&\psi_{A\uparrow}^\dagger(\textbf{R})\psi_{B\downarrow}^\dagger(\textbf{R}+\textbf{d}_i)
\\
&+\psi^\dagger_{B\uparrow}(\textbf{R}+\textbf{d}_i) \psi_{A\downarrow}^\dagger(\textbf{R})
\Big]+ hc\end{split}\nonumber\\
& &\hspace{-3mm}\begin{split}+\sum_{\textbf{R},\textbf{d}_i}\chi(\textbf{d}_i) \Big [&\psi_{A\uparrow}^\dagger(\textbf{R})\psi_{B\uparrow}(\textbf{R}+\textbf{d}_i)\\
 &+ \psi_{A\downarrow}^\dagger(\textbf{R})\psi_{B\downarrow}(\textbf{R}+\textbf{d}_i)\Big]+ hc\end{split}\nonumber \\
& &\hspace{-3mm}+\frac{1}{g_S}\sum_{\textbf{R},\textbf{d}_i}\Phi^\dagger(\textbf{R}+\textbf{d}_i) \Phi(\textbf{R}+\textbf{d}_i)
\nonumber \\
& &\hspace{-3mm}+\frac{1}{g_P}\sum_{\textbf{R},\textbf{d}_i\in \textbf{R}}\chi^\dagger(\textbf{R}+\textbf{d}_i)\chi(\textbf{R}+\textbf{d}_i) ,
\label{1a}
\end{eqnarray}
Varying with respect to $\Phi$ and $\chi$, we obtain, respectively,

\begin{eqnarray}
\Phi^\dagger(\textbf{R},\textbf{d}_i)= g_S\Big [\psi^\dagger_{A\uparrow}\psi_{B\downarrow}^\dagger
+ \psi_{B\uparrow}^\dagger
\psi^\dagger_{A\downarrow} \Big] 
\label{1b}
\end{eqnarray}
and
\begin{eqnarray}
\chi^\dagger (\textbf{R},\textbf{d}_i) =  g_P
\Big[\psi_{A\uparrow}^\dagger\psi_{B\uparrow}
+\psi^\dagger_{A\downarrow} \psi_{B\downarrow}
\Big]
\label{1c}
\end{eqnarray}

$\Phi^\dagger$ is a Cooper pair creation operator,  whereas $\chi^\dagger$ is an exciton creation operator. The vacuum expectation value of these operators, namely, $\Delta_\textbf{k}=\langle \Phi \rangle$, is a SC order parameter, while $M_\textbf{k}=\langle \chi \rangle$ is the PG order parameter. Cooper pair, as well as exciton formation occurs, respectively, for holes-holes or electron-holes, belonging to different sublattices. 

It follows from the perturbation theory structure (see for instance \cite{bj}) that
$$
\langle 0|\sum_\sigma \psi_{A\sigma}^\dagger\psi_{B\sigma}|0 \rangle = - \langle 0|\sum_\sigma \psi_{B\sigma}^\dagger\psi_{A\sigma}|0 \rangle.
$$
Hence we conclude that $M \equiv \langle \chi \rangle$ is a pure imaginary number: $M^*=-M$.

In momentum space, we have the corresponding Hamiltonian
\begin{eqnarray}
&\ &H_{eff}=\sum_{\textbf{k},\sigma}\epsilon(\textbf{k}) \Big [\psi_{A\sigma}^\dagger(\textbf{k})\psi_{B\sigma}(\textbf{k})+hc\Big ]
\nonumber \\
&\ &+ \sum_{\textbf{k}}\Phi(\textbf{k}) \Big [\psi_{A\uparrow}^\dagger(-\textbf{k})
\psi^\dagger_{B\downarrow}(\textbf{k}) + 
\psi^\dagger_{B\uparrow}(\textbf{k}) \psi_{A\downarrow}^\dagger(-\textbf{k})\Big]+hc
\nonumber \\
&\ &
+\sum_{\textbf{k}}\chi(\textbf{k}) \Big [\psi_{A\sigma}^\dagger(\textbf{k})
\psi_{B\sigma}(\textbf{k})  \Big ]+ hc
\nonumber \\
&\ &
%\hspace{-3mm}
+ \frac{1}{g_S}\sum_{\textbf{k}}\Phi^\dagger(\textbf{k}) \Phi(\textbf{k})
 \nonumber \\
&\ &+ \frac{1}{g_P}\sum_{\textbf{k}}\chi^\dagger(\textbf{k}) \chi(\textbf{k}),
\label{1aa}
\end{eqnarray}

where $\epsilon(\textbf{k})$ is the usual tight-binding energy, given by
\begin{eqnarray}
\epsilon(\textbf{k})=- t\sum_{i=1,...,4} e^{i \textbf{k}\cdot \textbf{d}_i}
\label{2x}
\end{eqnarray}\\
\bigskip \\
%\bigskip

{\bf 2.5) The d-Wave Character of the Order Parameters}\\
\bigskip

We want to derive an expression for the effective potential, which is a function of the ground-state expectation values: $\Delta$, $M$.  For this purpose, we use
the four-component Nambu fermion field, 

\begin{eqnarray}
\Psi_{a}=\left( \begin{array}{c}
\psi_{A,\uparrow,a} \\
\psi_{B,\uparrow,a}  \\
\psi^\dagger_{A,\downarrow,a} \\
\psi^\dagger_{B,\downarrow,a}
\end{array}               \right),
\label{2}
\end{eqnarray}
and replace the scalar fields with their ground-state expectation values. We then may rewrite
the hamiltonian in matrix form:
\begin{eqnarray}
&H_{eff} =& \frac{1}{g_S}\sum_{\textbf{k}}|\Delta(\textbf{k})|^2 +  \frac{1}{g_P}\sum_{\textbf{k}}|M(\textbf{k})|^2 \nonumber\\
& &+ \sum_{\textbf{k}} \Psi^{\dagger}_{a}(k) \mathcal{H}(k) \Psi_{a}(k) \label{Nambu}.
\end{eqnarray}

The index $a$ indicates to which of the parallel CuO$_2$ planes the electrons and holes belong and runs from 1 to $N$, where $N=1,2,3...$, according to the number of planes the specific material possesses.  In this approach, we shall neglect interplane interactions.

In the above expression
\begin{eqnarray}
\mathcal{H}
%\nonumber \\
=
\left(
\begin{array}{cccc}
0 & \epsilon + M & 0 &  \Delta
\\
\epsilon+ M^* & 0 & \Delta & 0
\\
0 & \Delta^{*} &0 & -\epsilon - M^*
\\
\Delta^{*} & 0 & -\epsilon - M & 0
\\
\end{array}
\right) 
\, .
\label{MatrizA1}
\end{eqnarray}

The energy eigenvalues are, then given by
 
\begin{eqnarray}
 E(\textbf{k})=\pm \sqrt{\epsilon^2(\textbf{k})+|M(\textbf{k})|^2 +|\Delta(\textbf{k})|^2}.
\label{eigenenergy}
\end{eqnarray}

Let us show here how the anisotropy in the  hybridization of the oxygen $p$-orbitals and the copper $d$-orbitals leads to the 
d-wave character of both the SC and PG order parameters.  

Firstly, notice that it follows from (\ref{1aa}) that
\begin{eqnarray}
& &\Delta(\textbf{k})= \sum_{i=1,..,4} \Delta(\textbf{d}_i)\exp\Big[i \textbf{k}\cdot \textbf{d}_i\Big ]
\nonumber \\
& &M(\textbf{k})= \sum_{i=1,..,4} M(\textbf{d}_i)\exp\Big[i \textbf{k}\cdot \textbf{d}_i\Big ]
\label{z1}
\end{eqnarray}

Then, notice that
$\Delta$ and $M$ in (\ref{MatrizA1}), effectively act as hopping parameters for the fermion field, similarly to the dimerization field in the Su-Schriefer-Heeger model for polyacetylene \cite{ssh,ecm2}. In that case, dimerization produces a nonzero ground-state expectation value of that field, which generates a gap for the electrons. In the case of the cuprates, the occurrence of nonzero values for $\Delta$ and $M$, respectively, produce a SC gap and the pseudogap. 

Now, observe that, because of the xy-anisotropy produced by the sign of the copper-oxygen orbital hybridization, as we can see in Fig. \ref{figa5}, we must have 
\begin{eqnarray}
\Delta(\textbf{d}_{1,3}) = - \Delta(\textbf{d}_{2,4})\equiv \Delta_0/2
\label{z2}
\end{eqnarray}
and
\begin{eqnarray}
M(\textbf{d}_{1,3}) = - M(\textbf{d}_{2,4})\equiv M_0/2
\label{z3}
\end{eqnarray}

It follows from (\ref{z1}) that

\begin{eqnarray}
& &\Delta(\textbf{k})
= \frac{\Delta_0 }{2}\left [e^{i \frac{(k_x + k_y)}{\sqrt{2}} a}+ e^{- i \frac{(k_x + k_y)}{\sqrt{2}} a}\right . 
\nonumber \\
& &\left .-e^{i \frac{(k_x - k_y)}{\sqrt{2}} a}- e^{- i \frac{(k_x - k_y)}{\sqrt{2}} a}\right ]
\label{d0}
\end{eqnarray}
hence 
\begin{eqnarray}
\Delta(\textbf{k})=\Delta_0 \left [\cos k_+ a - \cos k_-a \right ], 
\label{d1}
\end{eqnarray}
where
$$
k_{\pm}=\frac{k_x \pm  k_y}{\sqrt{2}}.
$$

Following precisely the same steps, as we did for $\Delta$, we may show that
\begin{eqnarray}
M(\textbf{k})=M_0 \left [\cos k_+ a - \cos k_-a \right ], 
\label{d2}
\end{eqnarray}
Also, from (\ref{2x}), we arrive at
\begin{eqnarray}
\epsilon(\textbf{k})= -2 t \left [\cos k_+ a + \cos k_-a \right ], 
\label{d3}
\end{eqnarray}

Using these expressions in (\ref{eigenenergy}),
we conclude that the above eigenvalues vanish at the four points $(k_x,k_y)=\textbf{K}= (\pm \pi/2, \pm \pi/2)$. Also we see that both the SC and PG oder parameters have lines of nodes along the $X$ and $Y$ directions, namely, along the directions where the copper ions are located. This  characterizes the d-wave nature of these parameters \cite{rmp1}.\\
\bigskip

{\bf 3) The Doping Process}

{\bf 3.1) The Mechanism of Doping}\\
\bigskip

The process of doping plays a central role in the physics of High-Tc cuprates. In this work, we will consider only hole doping, in which electrons are progressively removed from the oxygen $p_x$ and $p_y$ orbitals, thereby creating holes in such orbitals. The oxygen ions are themselves, placed on the sites of a square lattice, with a lattice parameter $a'=\sqrt{2}  a / 2 = 1.9\sqrt{2}$ \AA \, which possesses two sublattices, containing, respectively, $p_x$ and $p_y$ oxygen orbitals, which overlap with the $Cu^{++}$ d-orbitals (see Fig.\ref{fig01}), thereby forming bridges that will allow not only hole hopping along the whole oxygen lattice, but also the formation of Cooper pairs as well as excitons along these bridges. As we shall see, this fact naturally explains why both the SC and PG gaps have a d-wave symmetry.

In the case of the pure parent compounds the oxygen ions are doubly charged, namely:
$O^{--}$. Such ions are in a $2p^6$ configuration and the $p_x$ and $p_y$ orbitals contain two electrons each. The valence band which corresponds to the above described oxygen structure contains two electrons per site and, therefore, is completely filled. The electron density is $N_e=\frac{2}{A}$, where $A=a'^2$.
As doping is introduced, through some stoichiometric process, parametrized by $x$, one of the two electrons, either from the $p_x$ or the $p_y$ oxygen sublattices is pulled out of the plane, thereby creating a hole in such orbital. Expressing the average hole density per site in the oxygen lattice as $N_h=\frac{2}{A}y$, where $y\in [0,1]$, it follows that the average electron density becomes $N_e=\frac{2}{A}(1-y)$. Now, one must consider that the relation between the stoichiometric doping parameter, $x$ and the average number of holes per site in the oxygen lattice of the $CuO_2$ planes, associated to the $y$-parameter, is not universally known, in general; usually exhibiting different forms for each of the cuprate materials \cite{honma0,honma,honma1}. Consequently, we have the hole density parameter, $y$, given by some non-universal function of the doping parameter: $y=f(x)$. We typically do not know the function $f(x)$. In order to circumvent  this obstacle, we will describe the doping process through a constraint relating the fermion number directly to the stoichiometric doping parameter $x$, rather then to the density of holes in the oxygen lattice, which is parametrized by
$y$.

 Indeed, we write

\begin{eqnarray}
\lambda\Big[\sum_{a=1}^N \sum_{C=A,B} \psi^\dagger_{C,\sigma,a} \psi_{C,\sigma,a} - N d(x)\Big ]
\label{3}
\end{eqnarray}
where $d(x)$ is a function of the stoichiometric doping parameter, to be determined, and $N$ is the number of $CuO_2$planes. For consistency we must have $d(0)=\frac{2}{A}$, where $A=a'^2$ is the unit cell area of the oxygen lattice: $A=2\ (1.9)^2$ \ \AA$^2$. 

The constraint is enforced by integrating over the Lagrange multiplier field $\lambda$, whose vacuum expectation value is the chemical potential: $\langle \lambda \rangle = \mu$. It follows from (\ref{3}) that this must be proportional to $d(x)$. The proportionality constant between $\mu$ and $d(x)$ will be determined by fitting the experimental data

 As we increase the doping parameter $x$, the number of holes in the oxygen lattice will somehow increase as well, eventually reaching an amount where the critical SC temperature reaches a maximum. We call $x_0$ the value of the doping parameter for which this happens.
As we will see, the chemical potential vanishes precisely at the optimal doping $x=x_0$. 
\bigskip

{\bf 3.2) The Fermi Surface Formation}

The Fermi surface can be defined as the manifold for which the eigenvalues of $H-\mu\mathcal N$ vanish. Here $\mu=\mu(x)$ is the chemical potential of the holes and $\mathcal N=\mathcal N(x)$ is the hole number operator.

We have

\begin{eqnarray}
\mathcal{H}-\mu \mathcal{N} =
%\nonumber \\
\left(
\begin{array}{cccc}
-\mu & \epsilon + M & 0 &  \Delta
\\
\epsilon + M^* & -\mu & \Delta & 0
\\
0 & \Delta^{*} & \mu & -\epsilon -M^*
\\
\Delta^{*} & 0 & -\epsilon -M & \mu
\\
\end{array}
\right) 
\, .
\label{M2}
\end{eqnarray}

 The corresponding eigenvalues of $\mathcal H-\mu\mathcal N$,  are
\begin{eqnarray}
 \mathcal{E}(\textbf{k})=\pm \sqrt{(\sqrt{\epsilon^2(\textbf{k})+ |M(\textbf{k})|^2}\pm \mu)^2 +|\Delta(\textbf{k})|^2}.
\label{a1}
\end{eqnarray}

The Fermi surface, consequently, is defined by $ \mathcal{E}(\textbf{k})=0$. This leads to a second degree equation whose solution is

\begin{eqnarray}
 \mu(x)=\mp \sqrt{\epsilon^2(\textbf{k})+ |M(\textbf{k})|^2}\pm i |\Delta(\textbf{k})|
\label{a2}
\end{eqnarray}

Notice that the above expression becomes complex wherever a nonzero SC gap exists, reflecting the fact that no Fermi surface exists in the presence of a SC gap. 

We now consider the following regimes:

{\bf a) $T > T^*$}

In this case $M=\Delta=0$ and

\begin{eqnarray}
 \mu^2(x)=\epsilon^2(\textbf{k}) = v^2_{eff}\left ( \cos k_+a + cos k_-a \right)^2
\label{a3}
\end{eqnarray}

\begin{figure}
	[h]
	\centerline{
		%\figurename{TN}
		\includegraphics[scale=0.3]{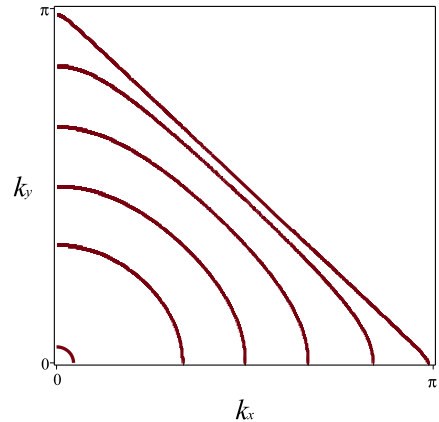}
		%clip, 
		%angle=90, 
		%width=1.\textwidth]
		%\include{TN}
	}
	\caption{Fermi surfaces for different levels of doping, at $T>T^*$ where $M=\Delta=0$. }
	\label{figa1}
\end{figure}

The corresponding Fermi surfaces are displayed in Fig. \ref{figa1}, for different values of the doping parameter.

{\bf b) $ T_c  < T < T^* $}

In this region, we have  $\Delta=0$ and $M \neq 0$. Then by making an expansion of $\epsilon(\textbf{k})$ and $M(\textbf{k})$, around the points \textbf{K}, where the energy eigenvalues are zero, and in terms of the  variables $k_\pm=\frac{k_x\pm k_y}{\sqrt{2}}$, we obtain

\begin{eqnarray}
&1& = \frac{ \left[ k_+ \mp \frac{\pi}{\sqrt{2}a} \right]^2}{\frac{\mu^2(x)}{2 v_{eff}^2}} + \frac{ k_-^2}{\frac{\mu^2(x)}{2[v_{\Delta}^2 + v_{M}^2]}} 
\nonumber \\
&1& = \frac{  k_+^2}{\frac{\mu^2(x)}{2 v_{eff}^2}} + \frac{\left[ k_{-} \mp \frac{\pi}{\sqrt{2}a}\right]^2}{\frac{\mu^2(x)}{2v_{M}^2}}
\end{eqnarray}
which are four ellipses centered at $(k_+,k_-)= (\pm \frac{\pi}{\sqrt{2}a}, 0)$ and $(k_+,k_-)=(0,\pm \frac{\pi}{\sqrt{2}a})$, with semi-axes given, respectively by $\frac{\mu(x)}{\sqrt{2} v_{eff}}$ and $\frac{\mu(x)}{\sqrt{2}v_{M}}$.

Equivalently, the ellipses will be centered at $(k_x,k_y)= (\pm \frac{\pi}{2a},\pm  \frac{\pi}{\sqrt{2}a})$
\begin{figure}. The independent curve will be

	[h]
	\centerline{
		%\figurename{TN}
		\includegraphics[scale=0.3]{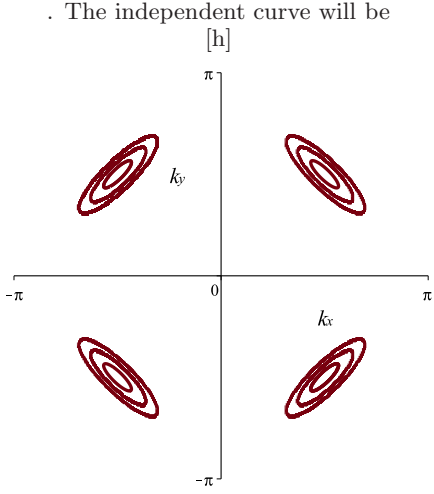}
		%clip, 
		%angle=90, 
		%width=1.\textwidth]
		%\include{TN}
	}
	\label{figa2}
	\caption{Fermi surfaces for different levels of doping at $T_c < T<T^*$ where $\Delta=0$ and $M \neq 0$. }
\end{figure}

These are the Fermi surfaces for the doped holes in the pseudogap region. 
Notice that the Fermi surface disappears whenever the chemical potential of the holes, $\mu(x)$, vanishes.  This occurs at zero stoichiometric doping parameter $x$, since it turns out that $\mu(x) \propto x$. For nonzero $x$,  the Fermi surface starts to show the pockets centered at the $\textbf{K}$ points.

This is precisely what is observed in ARPES experiments \cite{arpes}, thus puting our model in a solid experimental basis. 
Notice that all the pseudogap phenomenology, which is explained by the d-wave gap \cite{ddw} including the time-reversal, translation and rotation symmetries spontaneous breakdown, as well as the Nernst effect \cite{ddw1} are accounted for by our model as well.

As we have shown, the specific Hamiltonian interaction we use here, can be derived from a Spin-Fermion-Hubbard system, which describes the multiple magnetic interactions of a system of localized and itinerant spins, as well as the Coulomb repulsion between the itinerant degrees of freedom \cite{ecm1,ecm2}. A similar  Hamiltonian is described in \cite{vv}. Our elliptic constant energy curves would coincide with the ones obtained from an asymmetric kinetic Dirac lagrangean \cite{cms}, whereas the corresponding curves obtained from an usual Dirac lagrangean would correspond to circles. All of these must be in the same universality  class, therefore leading to the same phase diagram.

{\bf 4) The SC and PG Transition Temperatures: Derivation}\\
\bigskip

{\bf 4.1) Fermion  Integration}

We shall now integrate over the fermions, taking into account the doping constraint term, in order to obtain an effective potential in terms of the SC and PG order parameters and the chemical potential. The effective potential, despite being a function of the ground-state expectation values, actually takes into account fully quantized fluctuation effects \cite{ecm2}, being therefore a very useful tool for investigating the phase diagram of the system.

We can express the grand-canonical potential in terms of the effective potential as
\begin{eqnarray}
 \Omega [ \Delta,M,\mu  ]=\int d^2x d\tau V_{eff}\Big [ \Delta,M,\mu  \Big ],
\label{effpot1}
\end{eqnarray}
where
\begin{eqnarray}
& &\exp\Big \{ - \Omega [ \Delta,M,\mu  ]\Big \}=
\nonumber \\
& &\int D\Psi D\Psi^\dagger \exp\Big \{ \int d^2x d\tau \Big [ \frac{|\Delta|^2}{g_S} 
+\frac{|M|^2}{g_P} +N\mu d(x) 
\nonumber \\
& &+\Psi^\dagger \Big[ i \partial_\tau + \mathcal{H}[\Delta, M]-\mu \mathcal{N}\Big ] \Psi  \Big \}
\label{effpot}
\end{eqnarray}
where $ \mathcal{H}-\mu \mathcal{N} $ is given by (\ref{M2}).

Performing the quadratic functional integral over the fermion fields, after including the constraint term, we obtain the effective potential $V_{eff}[\Delta,M,\mu]$, namely,
\begin{eqnarray}
& &V_{eff}[\Delta, M,\mu] = \frac{|\Delta|^2}{g_S} 
+\frac{|M|^2}{g_P} +N\mu d(x) 
\nonumber \\
 & &+ N {\rm Tr} \ln \Big[i\partial_\tau +\mathcal{H}[\Delta, M]-\mu \mathcal{N} \Big ]
\label{2a}
\end{eqnarray}
Using the eigenvalues $\mathcal E(\textbf{k})$, given in (\ref{a1}), we can write

\begin{eqnarray}
&\hspace{-5mm}&V_{eff}[\Delta, M,\mu] = \frac{|\Delta|^2}{g_S} 
+\frac{|M|^2}{g_P} +N\mu d(x) 
\nonumber \\
& &\hspace{-5mm}- NT \sum_{n=-\infty}^\infty\sum_{l=\pm1}\int\frac{d^2k}{4\pi^2}
\nonumber \\
& &\hspace{-5mm}\ln\Big\{ \omega_n^2 +(\sqrt{\epsilon^2(\textbf{k})+| M(\textbf{k})|^2}+l \mu)^2 +|\Delta(\textbf{k})|^2 \Big\}\label{2aa}
\end{eqnarray}

Minimizing the effective potential with respect to the three variables, 
we find the following three equations:

\begin{eqnarray}
2\Delta_\textbf{k}\Big [-\frac{2T}{\alpha}F(\Delta_\textbf{k} , M_\textbf{k},\mu)+\frac{\eta(Ng_S)}{g_c}\Big ]=0
\label{7a}
\end{eqnarray}
\begin{eqnarray}
2M_\textbf{k}\Big [-\frac{2T}{\alpha}F(\Delta_\textbf{k} , M_\textbf{k},\mu)+\frac{\eta(Ng_P)}{g_c}\Big ]=0
\label{7b}
\end{eqnarray}
and
\begin{eqnarray}
d(x)=\mu \frac{4T}{\alpha}F(\Delta_\textbf{k} , M_\textbf{k},\mu)
\label{7c}
\end{eqnarray}
where $F(\Delta_\textbf{k} , M_\textbf{k},\mu)$ is a function, which, in the regime where $|\Delta_0|\sim 0 , |M_|0\sim 0$ is given by
\begin{eqnarray}
&\ &F(\Delta_0 , M_0,\mu_0)\Huge |_{|\Delta_0|\sim 0 , |M_0|\sim 0}=\ln2
\nonumber \\
&+&\frac{1}{2}\ln\cosh\Big[\frac{\sqrt{|\Delta_0|^2+(|M_0|+\mu_0(x))^2}}{2T}\Big]
\nonumber  \\
&+&\frac{1}{2}\ln\cosh\Big[\frac{\sqrt{|\Delta_0|^2+(|M_0|-\mu_0(x))^2}}{2T}\Big]
\nonumber \\
\label{7d}
\end{eqnarray}
and
\begin{eqnarray}
\eta(Ng)= \frac{Ng-g_c}{Ng}\ \ \ ;\ \ \  g_c=\frac{\alpha}{\Lambda}
\label{eta1}
\end{eqnarray}
%%%%%%%%%
Notice that $\eta(g)$ is a monotonically increasing function that saturates at infinity, namely 
\begin{eqnarray}
\eta(g) \stackrel{g\rightarrow \infty}\longrightarrow 1.
\label{eta2}
\end{eqnarray}
%%%%%%%%%%%%
In the expressions above, $\alpha=2\pi v_{eff}^2$, and $v_{eff}\simeq 2ta$ is the characteristic velocity and $\Lambda $  is an energy characteristic scale, which appears \cite{em} in connection to the characteristic length of the system. A natural choice is the coherence length $\xi$, which essentially measures the range of the pairing interaction (or the Cooper pair size). 
In cuprates we have $\xi\geq \xi_0 \simeq 10$\AA, whereas in conventional superconductors $\xi \geq\xi_0 \simeq 500$\AA.
 The energy cutoff is then  $\Lambda  \simeq 2\pi v_{eff} /\xi_0=\sqrt{2\pi\alpha}/\xi_0$. It determines the energy scale below which we may consider Cooper pairs as quasiparticles, hence it  must be of the order of $T_c$. 

We have 
\begin{eqnarray}
g_c=\frac{\alpha}{\Lambda} = \frac{\Lambda}{2\pi} \xi_0^2.
\label{gc}
\end{eqnarray}

We see that since $g_S\neq g_P$ it is impossible to satisfy (\ref{7a}) and (\ref{7b}) simultaneously with both $\Delta_0\neq 0$ and $M_0 \neq 0$, so we must have either $\Delta_0\neq 0$ and $M_0 = 0$  or $\Delta_0 = 0$ and $M_0 \neq 0$. The first is the SC phase, while the second is the PG phase.  

 For a fixed value of the doping parameter $x$, we have a gapless phase for $T>T^*$, the pseudogap phase, for $T_c <T <T^*$ whereas the superconducting phase sets in
at $T = T_c$.
As it turns out, the function $\frac{2T}{\alpha}F(\Delta, M,\mu)$ is monotonically decreasing, such that, for  $T>T^*$, we have $\frac{2T}{\alpha}F(\Delta_0, M_0,\mu) <\frac{\eta(Ng_P)}{g_c} < \frac{\eta(Ng_s)}{g_c} $ , thus implying, according to (\ref{7a})  and (\ref{7b}) that necessarily $\Delta_0, M_0 =0$. As we lower the temperature, we eventually reach  $T=T^*$, which characterizes the situation in which ${\alpha}F(\Delta_0, M_0,\mu) =\frac{\eta(Ng_P)}{g_c}$, hence, according to (\ref{7b}), we can have $M_0 \neq 0$ for  $T \leq T^*$. As we keep lowering the temperature, we eventually reach the situation where $\frac{2T}{\alpha}F(\Delta_0, M-0,\mu) $ grows enough to satisfy $\frac{2T}{\alpha}F(\Delta_0, M_0,\mu) = \frac{\eta(Ng_S)}{g_c}$, which, according to (\ref{7a}), implies we can have $\Delta_0 \neq 0$. 

For $T\leq T_c$, hence, we could have either $\Delta_0 \neq 0$ or $M_0 \neq 0$, however, as it turns out, the first condition is the energetically most favorable.\\
\bigskip

{\bf 4.2) The Superconducting Order Parameter }\\
\bigskip

Let us consider firstly the case $\Delta_0\neq 0$ and $M_0 = 0$.

Then (\ref{7a}) and (\ref{7c}) imply 
\begin{eqnarray}
\mu_0(x)=   d(x)\frac{g_c}{2\eta(Ng_S)} 
\label{11}
		\end{eqnarray}
where $g_c=\alpha/\Lambda$.

In order to find the critical temperature $T_c(x)$, we impose on (\ref{7a}) the condition $\Delta_0=0$ and $M_0=0$, which expresses the fact that the system is in one of the points belonging to the critical curve which separates the SC and PG phases.
Indeed, from (\ref{7a}), we obtain
\begin{eqnarray}
T_c(x)=\lim_{\Delta_0\rightarrow 0}\frac{ \frac{\alpha\eta(g_SN)}{2g_c}}{F(\Delta_0, M_0=0,\mu_0(x))}
\label{12}.
   \end{eqnarray}
 From this and (\ref{7d}), we see that, for $\Delta_0=0$ and $M_0=0$, the critical SC temperature, $T_c(x)$ satisfies
\begin{eqnarray}
T_c(x)=\frac{ \frac{\alpha\eta(g_SN)}{2g_c}}{\ln 2+\ln\cosh\Big[\frac{\mu_0(x)}{2T_c(x)}\Big]}.
\label{a12}
   \end{eqnarray}

It follows that
the upper bound of $T_c(x)$ occurs at a point $x=x_0$, where $\mu_0(x_0)=0$ and $T_{max}=T_c(x_0)$. Optimal doping occurs when the chemical potential vanishes. According to (\ref{11}), this implies $d(x_0)=0$. The simplest parametrization, for the case $N=1$, satisfying this and $d(0)=\frac{2}{A}$ is $d(x)=\frac{2}{Ax_0}(x_0-x)$, such that
\begin{equation}
 \mu_0=2\gamma(g_S)(x_0-x)
\label{mu},
\end{equation}
with
\begin{equation}
\gamma(g_S)=\frac{g_c}{2Ax_0 \eta(g_S)}
\label{gama}.
\end{equation}
The two equations above provide the link between the chemical potential and the stoichiometric doping parameter $x$. The parameter $\gamma$ will be determined by fitting the experimental data. Actually it will be the only parameter we fit.

Combining (\ref{mu}) with (\ref{12}) we can express the optimal temperature as
\begin{eqnarray}
T_{max}= \frac{\Lambda}{2\ln2} \eta(Ng_S)
\label{14a}
\end{eqnarray}
This should be compared with the corresponding BCS result, namely (see \cite{ecm2}, for instance)
\begin{eqnarray}
T_{c;BCS}= \frac{2 \gamma}{\pi}\hbar \omega_D e^{-\frac{1}{g_{BCS} N(E_F)}}
\label{14aa}
\end{eqnarray}
where $\gamma$ is the exponential of the Euler's constant $C\simeq 0.577$ and $\omega_D$ is the Debye frequency, a cutoff on the mediating phonon frequency, $g_{BCS}$ is the BCS coupling parameter and $N(E_F)$, the density of states at the Fermi level.

We can see that both our expression for the optimal $T_c$ in cuprates and the corresponding BCS result have a product of three similar factors: a universal numerical factor ($\frac{1}{2\ln2}$ in our case), a cutoff energy ($\Lambda$ in our case) and a function of the coupling parameter ($ \eta(Ng_S)$ in our case).

We see that our $T_{max}$ depends linearly on the inverse coupling whereas in conventional SC, there is an exponential dependence. 
The two functions, however, interestingly, are monotonically increasing functions of the coupling parameter that saturate at one as the coupling increases.

The linear dependence on the inverse coupling has been extensively studied before \cite{em,ecm2}. Notice that the function $\eta(x)$ is non-analytical in $x$, therefore indicating that our result is non-perturbative in the coupling parameter $g_S$, similarly to the BCS theory. 

By using the experimental values of $T_{max}$ for the many different compounds studied here, we find $\Lambda \simeq 0.018 eV$. This is compatible with values of $h v_{eff}$ and $\xi_0$, found in previous studies \cite{emsc}.

Inside the SC phase, we have $M_0=0$. Inserting this condition in (\ref{12}), we can derive an expression for the SC gap as a function of the temperature and doping, which is valid for $T\simeq Tc$  (note that both $\mu_0$ and $T_c$ depend on $x$)
\begin{eqnarray}
\cosh^2\Big(\frac{\sqrt{|\Delta_0|^2 +\mu^2_0}}{2T}\Big)=\cosh^2\Big(\frac{\mu_0}{2T_c} \Big)\exp\{2\ln2\}^{\frac{T_c}{T}-1}.
\label{12b}
\end{eqnarray}

Observe that $|\Delta_0(T_c,x)|=0$ and, since $\cosh$ is a monotonically increasing function, we must have
$|\Delta_0(T<T_c,x)|\neq 0$ for $T<T_c$. \\

The SC  gap at $T=0$ is given by
\begin{eqnarray}
|\Delta_0(T=0,x)|=\sqrt{\Big[2\ln 2 T_c(x)\Big]^2-\mu_0(x)^2},
\label{12d}
   \end{eqnarray}
from which we obtain the following ratio between the optimal SC critical temperature and the zero temperature gap at the optimal doping ($x=x_0$):
\begin{eqnarray}
\frac{|\Delta_0(T=0,x_0)|}{T_{max}} =2\ln 2 .
\label{12e}
 \end{eqnarray}  
This is an universal ratio, which apparently applies to all cuprates.\\
\bigskip

{\bf 4.3) The Pseudogap Order Parameter }\\
\bigskip

We consider now the case where  $\Delta_0 = 0$ and $M_0 \neq 0$. In order to find the critical temperature $T^*(x)$, we take (\ref{7b}) in the limit $M_0\rightarrow 0$, which leads to
\begin{eqnarray}
T^*(x)=\lim_{M_0\rightarrow 0}\frac{ \frac{\alpha\eta(Ng_P)}{2g_c}}{F(\Delta_0=0, M_0,\tilde\mu_0(x))}
\label{12bb}.
   \end{eqnarray}

Now (\ref{7c}) yields the following expression for the chemical potential
\begin{equation}
\tilde\mu_0=2\tilde\gamma(Ng_P)(\tilde x_0-x).
\end{equation}
Observe that, because $M_0\neq 0$ in the PG phase, the chemical potential $\tilde\mu_0(x)$, no longer  vanishes at the optimal doping $x_0$. \\

Inside the PG phase, we have $\Delta_0=0$. Inserting this condition in (\ref{7b}), we can derive an expression for determining the PG gap as a function of the temperature and doping, which is valid for $T\simeq T^*$.
%\begin{eqnarray}
%\sinh^2\Big(\frac{M_0}{2T}\Big)= \exp\{2\ln2\}^{\frac{\tilde T}{T}-1} - \cosh^2\Big( \frac{\tilde\mu_0(x)}{2T}\Big)
%\label{12b1}
%\end{eqnarray}
%or, equivalently
\begin{eqnarray}
&\sinh^2\Big(\frac{M_0}{2T}\Big)&=  \Big [\cosh^2\Big( \frac{\tilde\mu_0(x)}{2T^*}\Big) \Big ]^{\frac{T^*}{T}} \exp\{2\ln2\}^{\frac{T^*}{T}-1}
\nonumber \\
& & -\cosh^2\Big( \frac{\tilde\mu_0(x)}{2T}\Big)
\nonumber \\
\label{12b1}
\end{eqnarray}\\
\bigskip

{\bf 4.4) The Critical SC Temperature: $T_c(x)$ }\\
\bigskip

The critical curve delimiting the boundary of the SC phase is obtained from (\ref{a12}), however, we must be careful when taking the limit 
\begin{eqnarray}
\frac{T_c(x)}{\ln2 T_{max}} &=&\lim_{|M_0|\rightarrow 0} \Big\{\ln2+\frac{1}{2}\ln\cosh\Big[\frac{||M_0|+\mu_0(x)|}{2T_c(x)}\Big]
\nonumber  \\
&+&\frac{1}{2}\ln\cosh\Big[\frac{||M_0|-\mu_0(x)|}{2T_c(x)}\Big]\Big\}^{-1}.
\label{12a}
   \end{eqnarray}
Indeed,  considering that $\mu_0(x)$ has different signs for $x_0 -x>0$ and $x_0 -x < 0$, we arrive at different equations for $T_c(x)$ in the underdoped, $x<x_0$ and overdoped, $x>x_0$ regions.

By putting in evidence the first exponential from the hyperbolic cosine, we obtain 
\begin{eqnarray}
T_c(x) =\lim_{|M_0|\rightarrow 0}\frac{\ln2 \ \ T_{max}}{B(x) +\ln2+\frac{1}{2}\Big \{ \exp\left[ - \frac{2\gamma (x_0-x)}{T_c(x)} \right] - 1  \Big \}},
\nonumber \\
\label{18}
\end{eqnarray}
where,
$$
B(x)=\lim_{|M_0|\rightarrow 0}\left[\frac{|\mu_0(x)+|M_0||}{4T_c(x)}+\frac{|\mu_0(x)-|M_0||}{4T_c(x)}\right]
$$
Now, for $x<x_0$, we have $\mu_0=2\gamma(x_0-x) >0$ and, consequently, 
\begin{eqnarray}
& &B(x)=\lim_{|M_0|\rightarrow 0}\left[\frac{\mu_0(x)+|M_0|}{4T_c(x)}+\frac{\mu_0(x)-|M_0|}{4T_c(x)}\right ]
\nonumber \\
& &=\frac{\mu_0(x)}{2T_c(x)}
\label{b18}
\end{eqnarray}
and 
\begin{eqnarray}
T_{c}(x) =\frac{\ln2 \,\ T_{max}}{\ln2 + \frac{\mu_0(x)}{2T_{c}(x)} + \frac{1}{2}\left(e^{-\frac{\mu_0(x)}{T_{c}(x)}}-1\right)}
\, ,
\hspace{0.3cm}
x < x_{0}
\, .
\label{ud}
\end{eqnarray}

For $x>x_0$, , however, $\mu_0 <0$,and we have
\begin{eqnarray}
& &B(x)=\lim_{|M_0|\rightarrow 0}\left [\frac{|M_0|-|\mu_0(x)|}{4T_c(x)}+\frac{|M_0|+|\mu_0(x)|}{4T_c(x)}\right ]
\nonumber \\
& &=\lim_{|M_0|\rightarrow 0}\frac{|M_0|}{2T_c(x)}=0
\label{c18}
\end{eqnarray}

and $T_c(x)$ is given by
\begin{eqnarray}
T_c(x) =\frac{\ln2 \ \ T_{max}}{\ln\Big [1+ \exp\left[- \frac{\mu_0(x)}{T_c(x)} \right]  \Big ]}
\, ,
\hspace{0.5cm} x > x_{0}
\, .
\label{od}
		\end{eqnarray}

From the above expressions for $T_c(x)$, we may determine  the two quantum critical points,$ x^{\pm}_{SC}$, where the SC dome starts at $T=0$
Taking the limit $T_c \rightarrow 0$ in (\ref{ud}) and (\ref{od}), we find
\begin{eqnarray}
x^{-}_{SC} = x_0 - \frac{T_{max}}{\gamma}\ln 2 \ \ \ ;\ \ \ x^{+}_{SC} = x_0 + \frac{T_{max}}{2\gamma}\ln 2
\label{x17}
		\end{eqnarray}

We see that the SC dome is, in general asymmetric with respect to the optimal doping, a feature that is corroborated by many experimental data. LSCO is, apparently the only exception \cite{honma,honma1,honma0} and for it, we have 

\begin{eqnarray}
T_{c}(x) =\frac{\ln2 \,\ T_{max}}{\ln2 + \frac{|\mu_0(x)|}{2T_{c}(x)} + \frac{1}{2}\left(e^{-\frac{|\mu_0(x)|}{T_{c}(x)}}-1\right)}
\label{udLSCO}
\end{eqnarray}
both in the underdoped ($x<x_0$) and overdoped ($x>x_0$) regions.

For LSCO, the quantum critical points  $x^{\pm}_{SC}$ are symmetric about $x_0$, namely,
\begin{eqnarray}
x^{\pm}_{SC} = x_0 \pm \frac{T_{max}}{\gamma}\ln 2 .
\label{xx17}
		\end{eqnarray}\\
\bigskip

{\bf 4.5) The Critical PG Temperature: $T^*(x)$}\\
\bigskip

We are now going to obtain the critical line delimiting the PG phase, namely  $T^*(x)$. For this purpose, we start from (\ref{12bb}) and taking the limit $M_0\rightarrow 0$, obtain
\begin{eqnarray}
T^*(x) =\frac{\frac{\alpha\eta(g_PN)}{2g_c}}{\ln\Big [1+ \exp\left[- \frac{2\tilde\gamma (\tilde x_0 - x)}{T^*(x)} \right]  \Big ]}.
\label{19a}
		\end{eqnarray}
%\\
%\bigskip
Observe, however, that now we are on the solution of (\ref{7b}), instead of (\ref{7a}), hence we must replace $g_S$ with $g_P$ and, accordingly, $\eta$ and $\gamma$ with $\tilde\eta$ and $\tilde\gamma$, which satisfy, for the case $N=1$,
\begin{eqnarray}
\tilde\gamma \tilde x_0 \tilde \eta =\gamma x_0 \eta
\label{19y}
		\end{eqnarray}
From (\ref{19a}), we see that $\tilde x_0$ is the point where the function $T^*(x)\rightarrow 0$. We therefore have $\tilde x_0= x^+_{SC}$, where the latter is given by (\ref{x17}).\\
\bigskip

{\bf 5) The SC and PG Transition Temperatures: Applications}\\
\bigskip

{\bf 5.1) Determination of Parameters}\\
\bigskip

We present a summary of the relevant parameters for each of the compounds considered in this study in Table \ref{t1}.

For determining the values of the relevant parameters for each family group (Bismuth family, Mercury family, LSCO)  we proceed through the following steps:

a)
For a set of compounds consisting of $N$ materials, all belonging to the same family, each of them possessing  $m=1,...,N$ $CuO_2$ planes, we must determine firstly the $N$ parameters: $\eta(mg)$, for  $m=1,...,N$. 
For this purpose, notice that
from (\ref{eta1}) considered at the values: $m=2,...,N$, we have $N-1$ linear equations relating the $\eta(mg)$'s, namely,
\begin{eqnarray}
\eta(mg)= \frac{m-1}{m} +\frac{1}{m}\eta(1g).
\label{eta3}
\end{eqnarray}
Notice that, in order to obtain the above equation, we must assume the coupling parameters, either $g=g_S$ (or $g=g_P$, below) in $\eta(Ng)$ are the same for all compounds which are members of the same family.
 We then combine the $N-1$ linear equations (\ref{eta3}), with
\begin{eqnarray}
 \frac{T_{max}(N=2)}{T_{max}(N=1)} =\frac{\eta (2g)}{\eta (1g)} ,
\label{eta4}
\end{eqnarray}
thus obtaining a set of $N$ linear equations relating the $N$ parameters $\eta(mg)$, for  $m=1,...,N$. This would allow us to determine the $\eta(mg)$ parameters for any $N$, $m=1,...,N$. 

In Section 6, below, we argue that equation (\ref{eta4}) will be experimentally accurate up to ratios of $\frac{T_{max}(N=3)}{T_{max}(N=1)}$, hence it can be safely used for $N=2$.

b) For the $Bi$ and $Hg$ families, first find $\eta(n=1)$, $\eta(n=2)$ and $\eta(n=3)$ by using
\begin{eqnarray}
&1)&     \frac{T_{max}(n=2)}{T_{max}(n=1)} =\frac{\eta (n=2)}{\eta (n=1)} \nonumber \\    \nonumber \\
&2)& \eta (n=3)=\frac{\eta (n=1)}{3} +\frac{2}{3}\nonumber \\   \nonumber \\
&3)&\eta (n=2)=\frac{\eta (n=1)}{2} +\frac{1}{2}
\label{w19w}
\end{eqnarray}
where the last two equations derive from  (\ref{eta4}), assuming that the coupling parameters $g_S$ are the same for all compounds of the same family.

From the three equations above, using the experimental values of $T_{max}(n=2)$ and $T_{max}(n=1)$, we 
determine $\eta(n=1) = 0.23077$, $\eta(n=2) = 0.61538$, $\eta(n=3) = 0.74358$ for the $Bi$ family and $\eta(n=1) = 0.61577$, $\eta(n=2) = 0.80788$, $\eta(n=3) = 0.87192$ for the $Hg$ family;

c) Now use 
\begin{eqnarray}
T_{max}(n)=\frac{\eta (n)}{2\ln 2} \Lambda
\label{19www}
\end{eqnarray}
to determine $\Lambda$. We find for all the materials of the $Bi$ and $Hg$ families: $\Lambda = 0.018 eV$.

d)  for LSCO, use the value of $\Lambda$ found above and (\ref{14a}) to obtain $\eta = 0.23846$;

e) Obtain the ratio $g_S/g_c$ from
\begin{eqnarray}
\frac{Ng_S}{g_c}=\frac{1}{1-\eta(N)};
\label{w18}
\end{eqnarray}

f) $g_c=\frac{\alpha}{\Lambda}$ is determined by inserting in (\ref{gc}) the value of the characteristic energy scale $\Lambda=0.018$ eV, which was determined above and of the corresponding characteristic length scale $\xi_0 \simeq 10$ \AA, which is known experimentally. We find $g_c = 0.30$ eV.

g) From e and f, determine $g_S$;

h) Adjust only $\gamma$ for the curves $T_c(x)$ and only $\tilde\gamma$ for the curves $T^*(x)$ to fit the experimental data; then, from (\ref{19y}) find $\tilde\eta(N=1)$;

i) Use (\ref{eta3}) to determine $\tilde\eta(n)$ from $\tilde\eta(n=1)$;

j) Determine $g_P$ from 
\begin{eqnarray}
\frac{Ng_P}{g_c}=\frac{1}{1-\tilde\eta(N)};
\label{ww18}
\end{eqnarray}
and from $g_c$, found in item d.

A summary of the results found by following the steps described above is shown in Table \ref{t1}.
\bigskip 
{\bf 5.2) The SC  Critical Temperatures}\\
\bigskip

We now apply the previous results to several High-Tc cuprates, namely, the one-layered, LSCO, Bi2201 and Hg1201, the two-layered Bi2212 and Hg1212 and the three-layered Bi2223 and Hg1223. We use MAPLE in order to obtain the curves $T_c(x)$ and $T^*(x)$, satisfying
(\ref{ud}), (\ref{od}).

For obtaining $T_c(x)$, we enter the experimental values of $T_{max}$ and $x_0$ and adjust only one parameter, namely $\gamma$, for the curve $T_c(x)$ to fit the experimental data.

\begin{table}[!ht]
\begin{tabular}{|c|c|c|c|c|c|c|}
\hline\hline
           & N & $T_{max}$ (eV) &   $x_0$   & $\gamma$ (eV) & $\eta$ & $g_S$ (eV) \\ \hline \hline
Bi2201 & 1    &  0.0030         &    0.29       &      0.012 & 0.23077 & 0.39000       \\ %\hline
Bi2212  & 2  &  0.0080        &   0.245       &   0.041  & 0.61538  &  0.39000\\% \hline
Bi2223   & 3  &   0.0093        &    0.212      &    0.049 & 0.74358 &0.39000  \\ \hline
Hg1201 & 1   &   0.00835        &  0.25        &    0.031 &0.61577 &0.7818 \\% \hline
Hg1212 & 2  &    0.0111       &   0.24       &      0.044  & 0.80788 &  0.7818 \\ %\hline T_{max}=0.0108
Hg1223 & 3  &  0.0115         &   0.214       &     0.054  & 0.87192  & 0.7818 \\ \hline
LSCO    & 1  &   0.0031        &     0.16       &   0.020   &   0.23870   & 0.39406\\ \hline
\end{tabular}
\caption{The parameters used for obtaining the $T_c(x)$ curves. Only $\gamma$ has been adjusted. The last column displays the value obtained for the coupling parameter $g_S$.}
\label{t1}
\end{table}

\begin{figure}[!ht]
	\centering
	\includegraphics[scale=0.30]{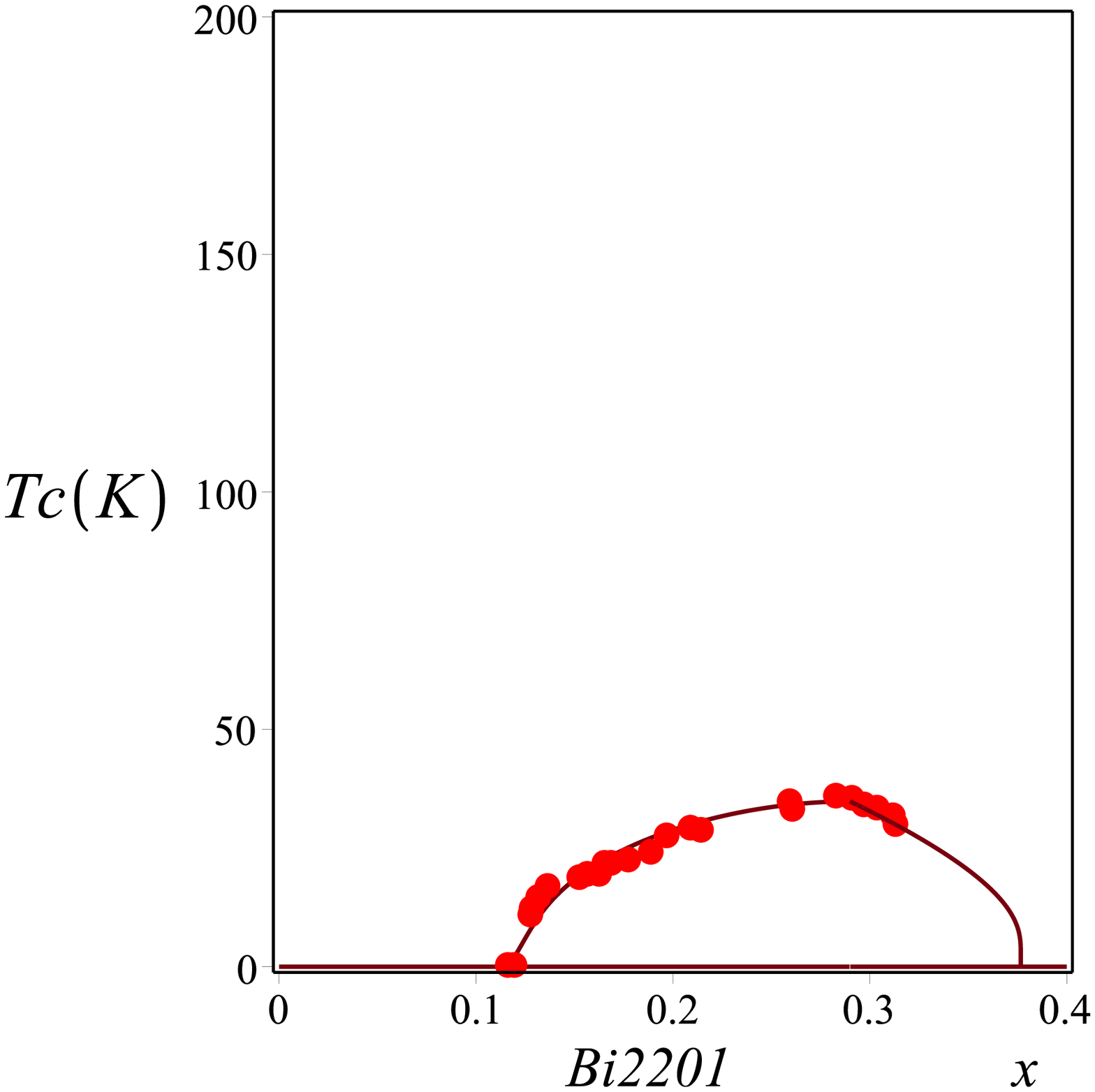}
	\caption{Solution of Eqs. (\ref{ud}) and (\ref{od}) for the SC dome of Bi2201. Experimental data from \cite{5,6}.} \label{fig2}
\end{figure}

%%%%%%%%%%%%%%%%%%%%%%%%%%%%%%

\begin{figure}[!ht]
	\centering
	\includegraphics[scale=0.30]{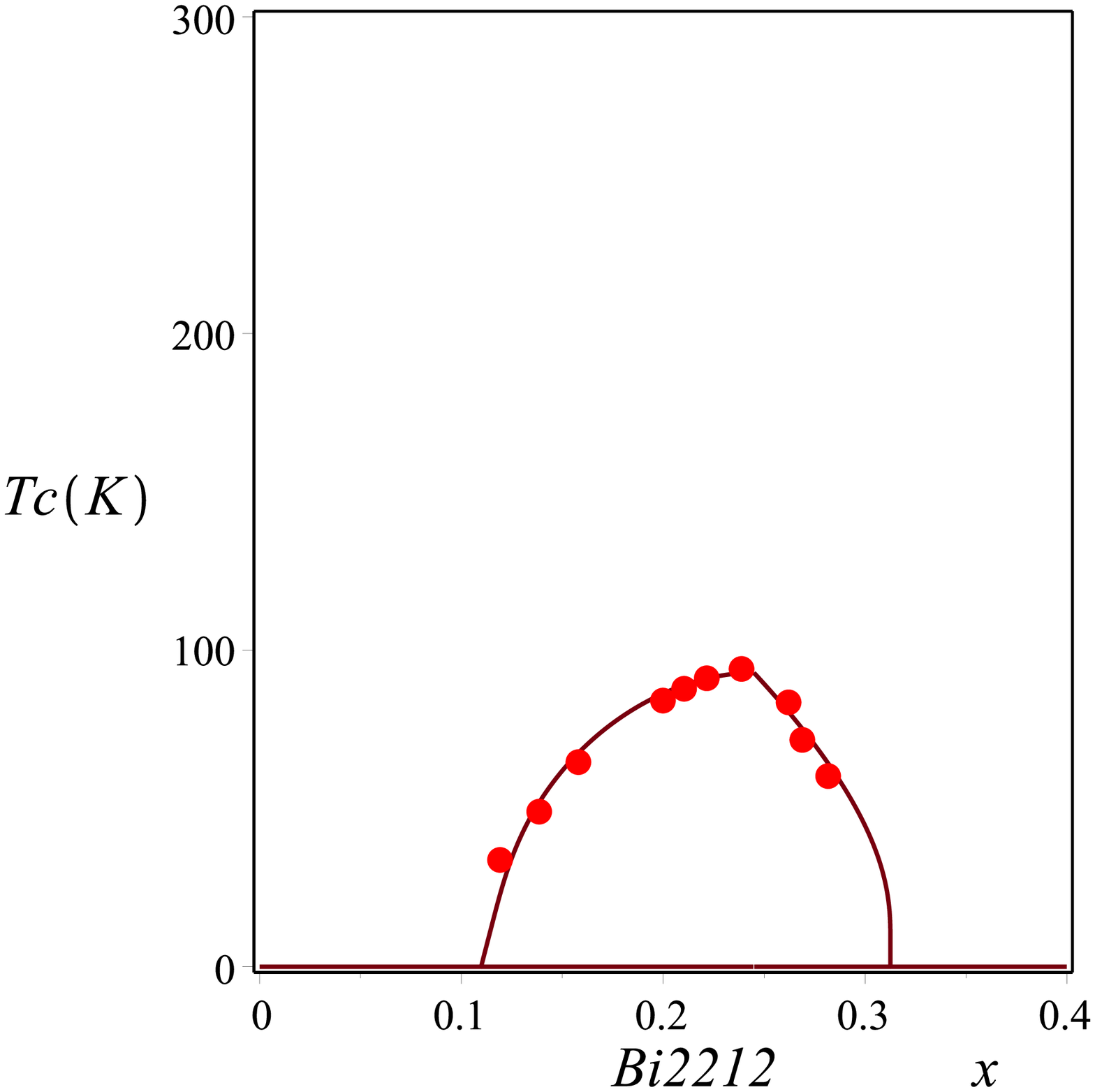}
	\caption{Solution of Eqs. (\ref{ud}) and (\ref{od}) for the SC dome of Bi2212. Experimental data from \cite{7,8}.} \label{fig4}
\end{figure}

\begin{figure}[!ht]
	\centering
	\includegraphics[scale=0.30]{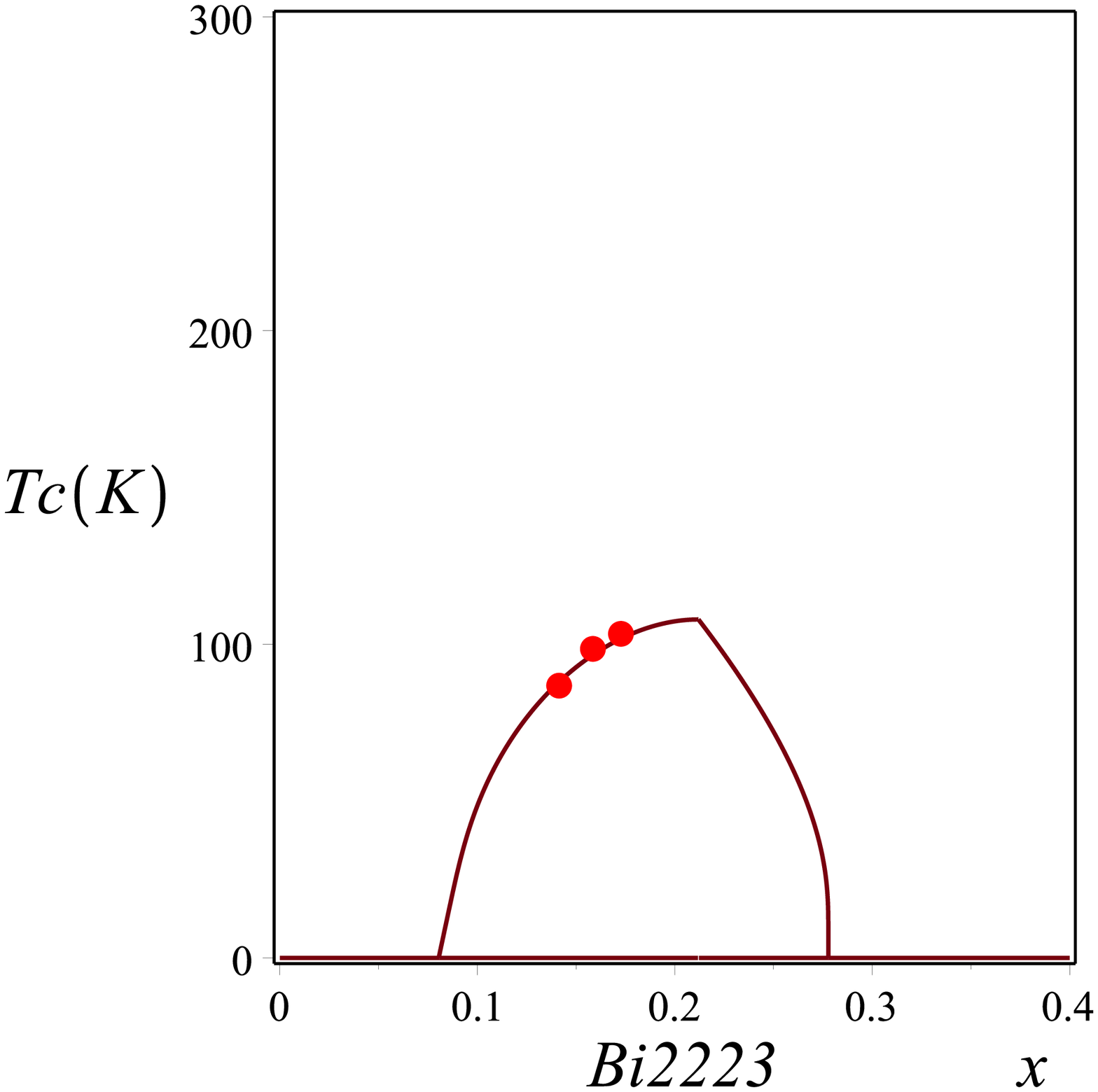}
	\caption{Solution of Eqs. (\ref{ud}) and (\ref{od}) for the SC dome of Bi2223. Experimental data from \cite{7,8}.} \label{fig5}
\end{figure}

\begin{figure}[!ht]
	\centering
	\includegraphics[scale=0.30]{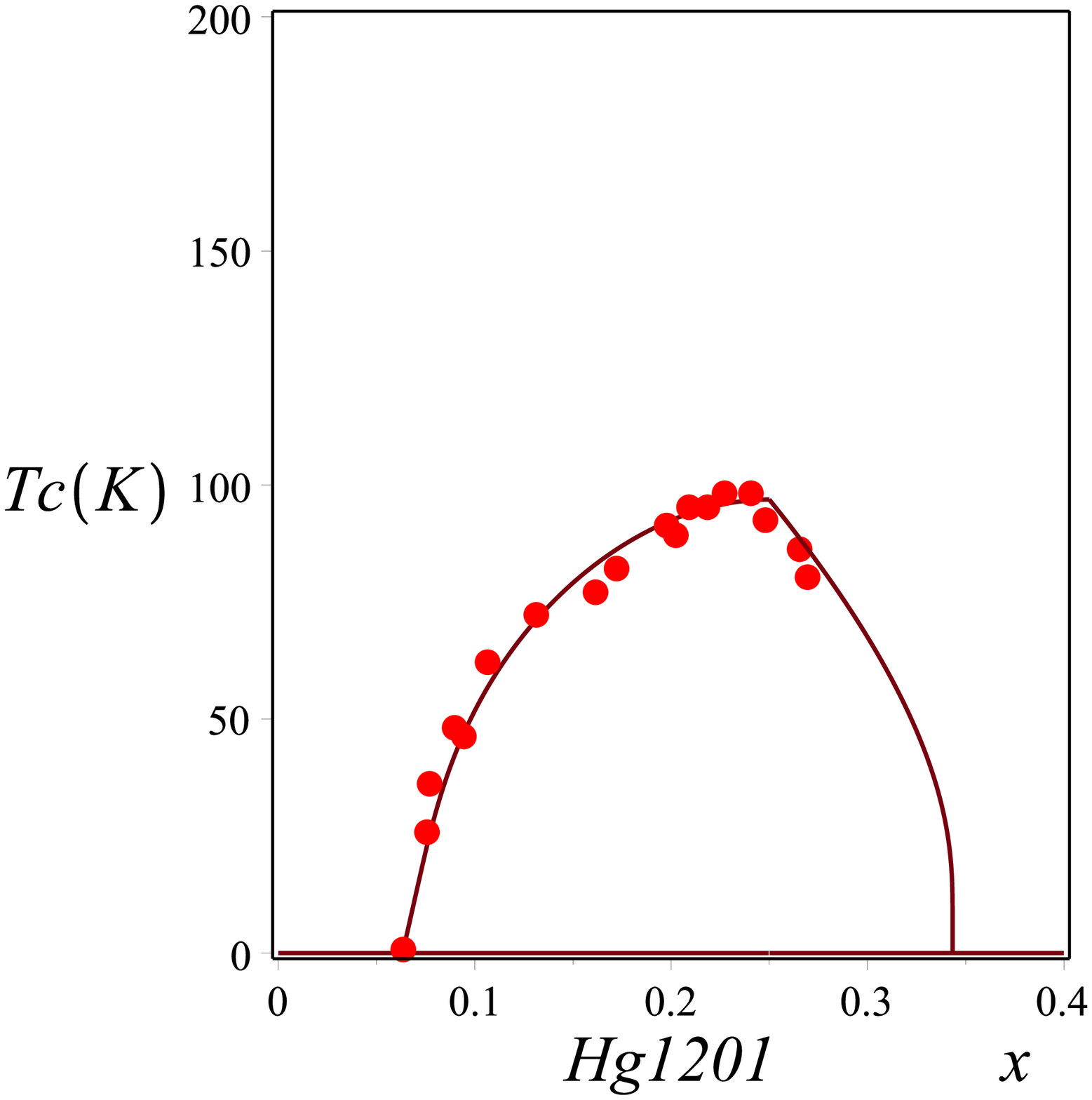}
	\caption{Solution of Eqs. (\ref{ud}) and (\ref{od}) for the SC dome of Hg1201. Experimental data from \cite{7,8}.} \label{fig3}
\end{figure}

\begin{figure}[!ht]
	\centering
	\includegraphics[scale=0.30]{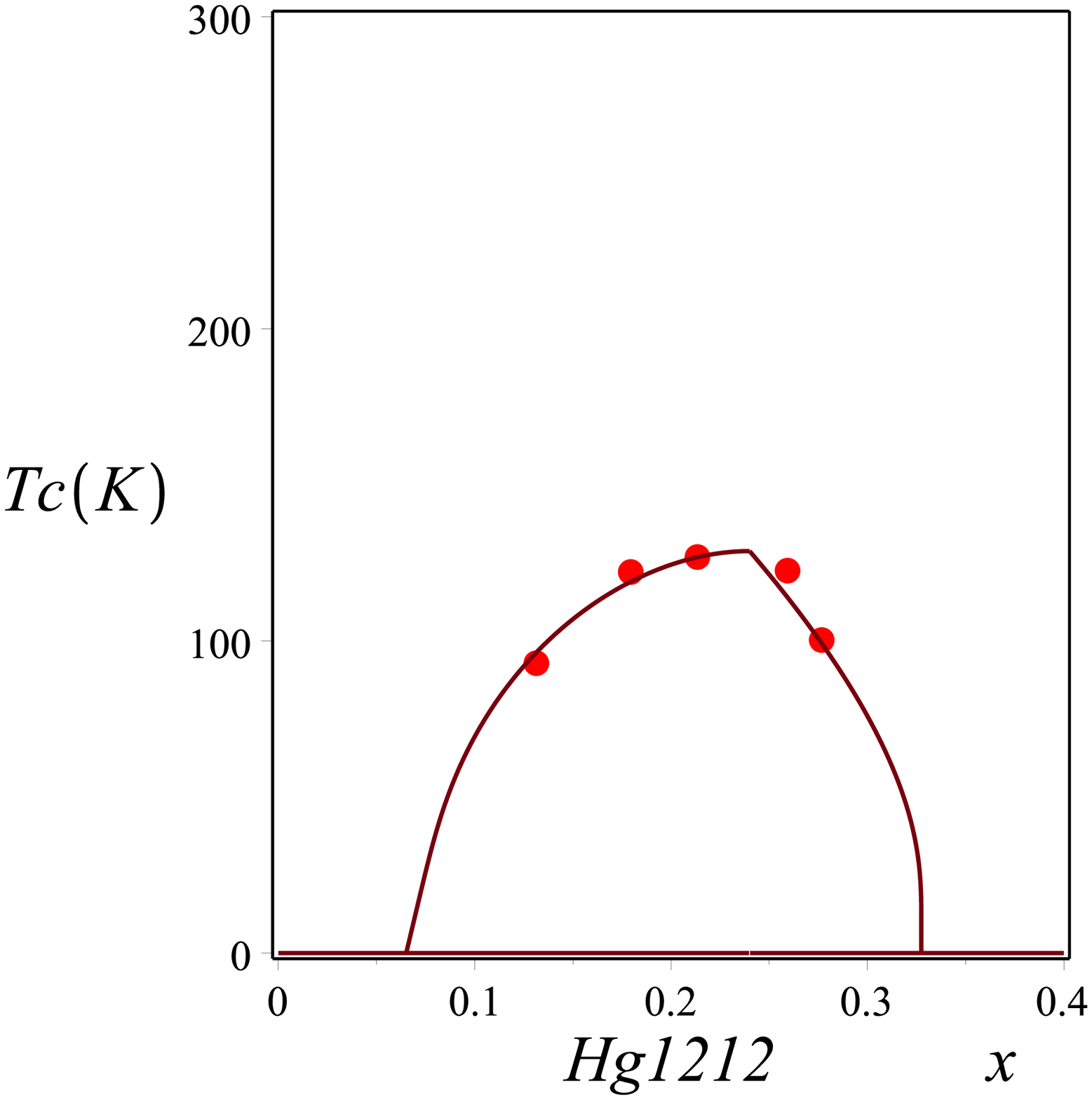}
	\caption{Solution of Eqs. (\ref{ud}) and (\ref{od}) for the SC dome of Hg1212. Experimental data from \cite{7,8}.} \label{fig6}
\end{figure}

\begin{figure}[!ht]
	\centering
	\includegraphics[scale=0.30]{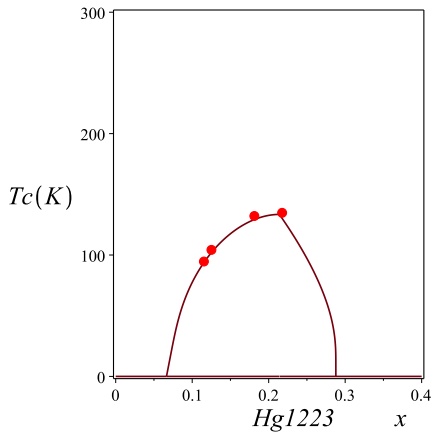}
	\caption{Solution of Eqs. (\ref{ud}) and (\ref{od}) for the SC dome of Hg1223. Experimental data from \cite{7,8}.} \label{fig7}
\end{figure}
\bigskip

{\bf 5.3) The PG  Critical Temperatures}\\
\bigskip

We now present the results for the curves $T_c(x)$ and $T^*(x)$, satisfying (\ref{ud}), (\ref{od}) and (\ref{19a}), for the same materials considered above.

In order to fit the curve $T^*(x)$ to the experimental data, again we adjust only one parameter, namely $\tilde\gamma$, given by
\begin{eqnarray}
\tilde\gamma=\frac{g_c}{2A \tilde x_0 \tilde\eta}.
\label{tg}
		\end{eqnarray}
The parameter $\tilde x_0$ in (\ref{tg}) coincides with  $x^+_{SC}$, since it is the point where $T^*(x)$ vanishes in (\ref{19a}).

We summarize in Table \ref{t2} the parameters related to the curves $T^*(x)$:

\begin{table}[]
\begin{tabular}{|c|c|c|c|c|}
\hline\hline
           & $\tilde\gamma$ & $\tilde x_0$ &  $\tilde\eta$ (eV)  & $g_S/g_P$\\ \hline \hline
Bi2201 & 0.1320 &  0.376        &              0.01618                          & 1.28        \\ %\hline
Bi2212  & 0.2708  &  0.32       &             0.50809                           &  1.28 \\ %\hline
Bi2223   & 0.1100  &   0.25        &           0.67205                          & 1.28  \\ \hline
Hg1201 & 0.1860   &   0.343        &           0.07480                         &   2.50 \\% \hline
Hg1212 & 0.0890  &    0.29          &           0.53740                         & 2.50 \\% \hline
Hg1223 & 0.2200  &  0.247         &             0.69159                         &2.50  \\ \hline
LSCO    & 0.1800  &   0.267        &            0.01565                          & 1.29 \\ \hline
\end{tabular}
\caption{Relevant parameters for different cuprates. It was assumed the couplings $g_S$ and $g_P$ are the same for all members of a family.}
\label{t2}
\end{table}

On the last column, we list the ratio $\frac{g_S}{g_P}$, given by (\ref{w18}).

\begin{figure}[!ht]
	\centering
	\includegraphics[scale=0.30]{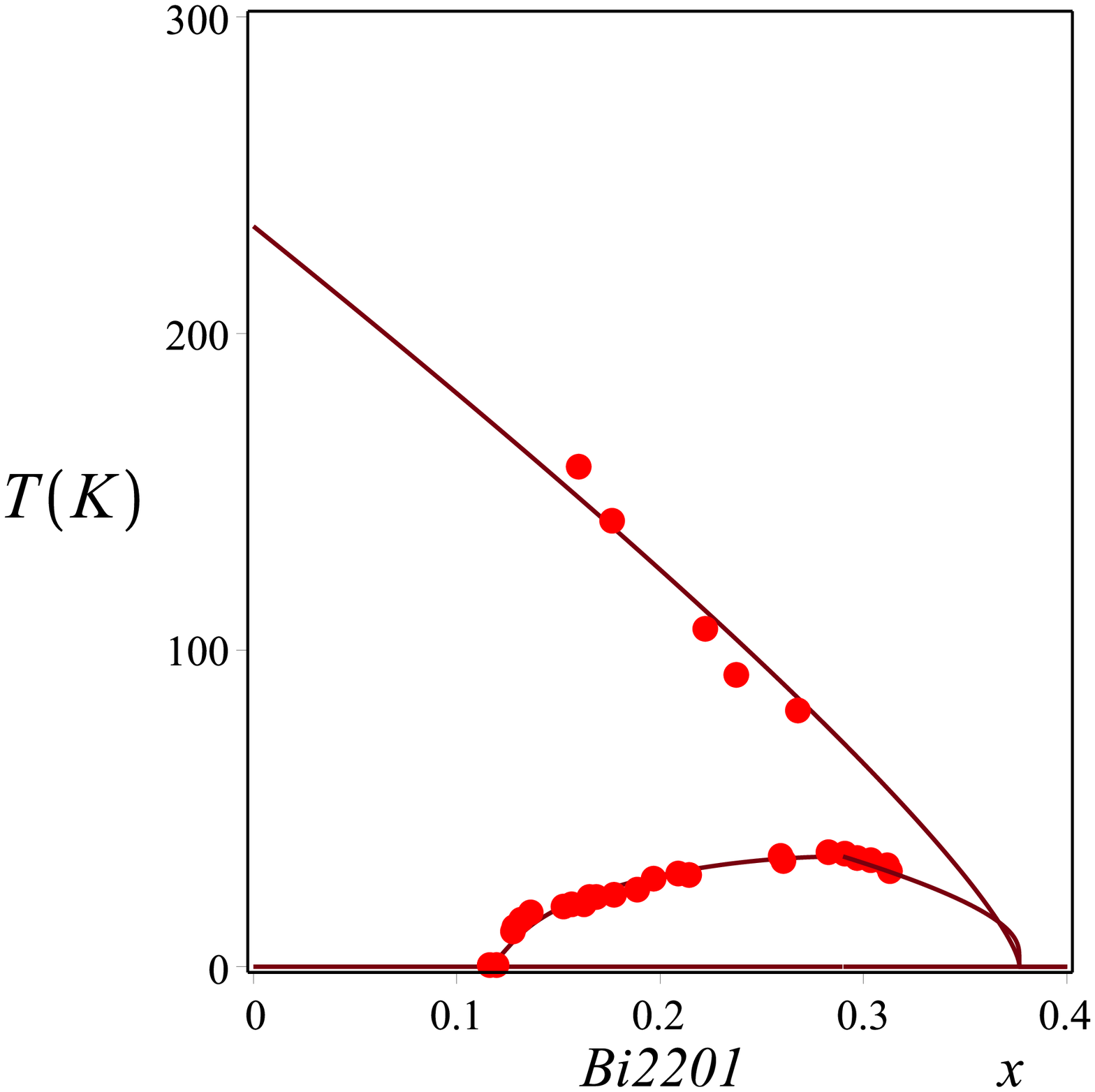}
	\caption{Solution of Eqs. (\ref{ud}) and (\ref{od})  for the SC dome of Bi2201, together with the solution of (\ref{19a}) for the pseudogap temperature $T^*(x)$. Experimental data for $T_c(x)$ from \cite{1,2,3,4} and for $T^*(x)$ from \cite{7}.} \label{figb}
\end{figure}

\begin{figure}[!ht]
	\centering
	\includegraphics[scale=0.30]{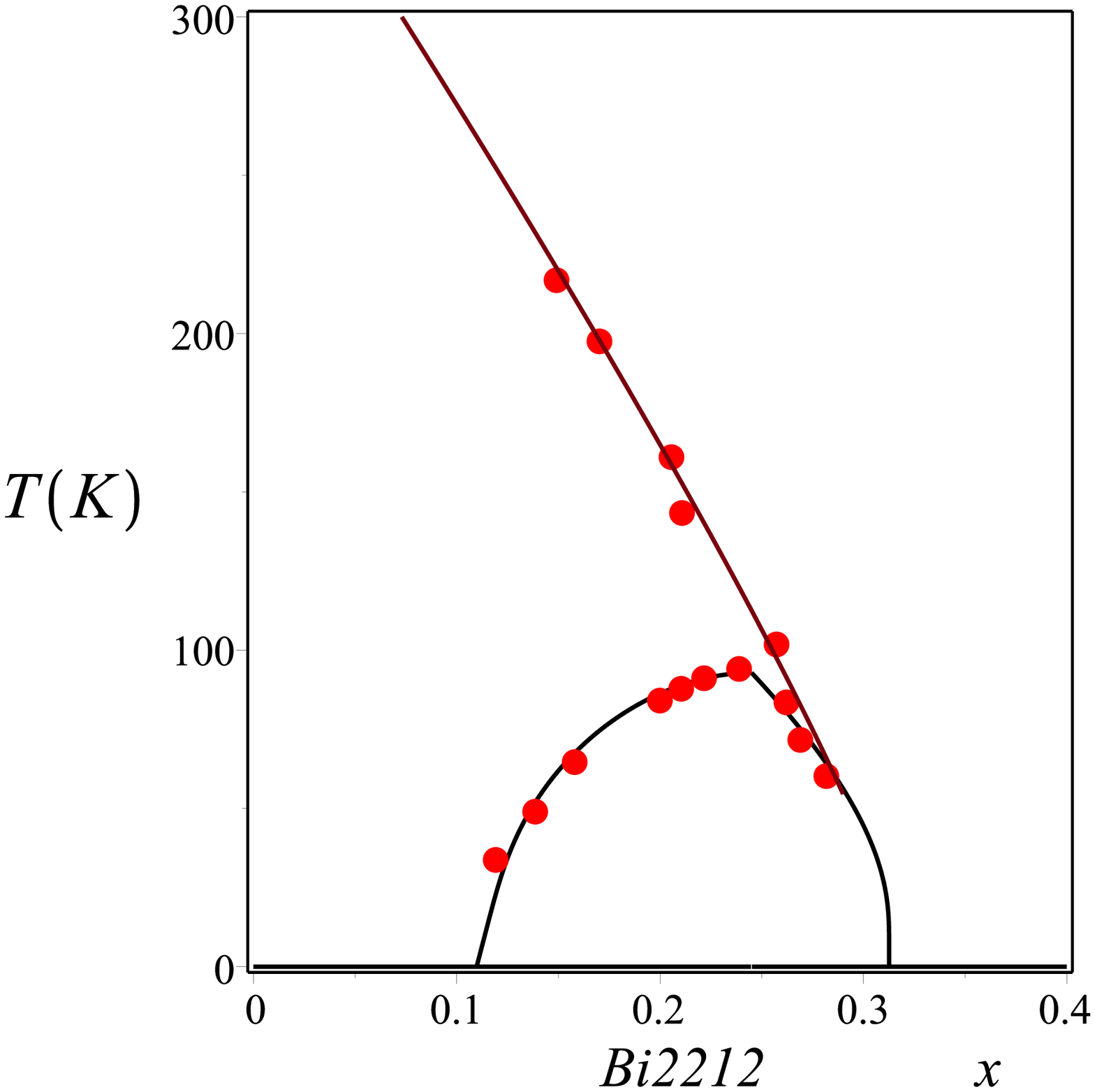}
	\caption{Solution of  Eqs. (\ref{ud}) and (\ref{od})  for the SC dome of Bi2212, together with the solution of (\ref{19a}) for the pseudogap temperature $T^*(x)$. Experimental data for $T_c(x)$ and for $T^*(x)$ both from\cite{7,8}.} \label{figd}
\end{figure}

\begin{figure}[!ht]
	\centering
	\includegraphics[scale=0.30]{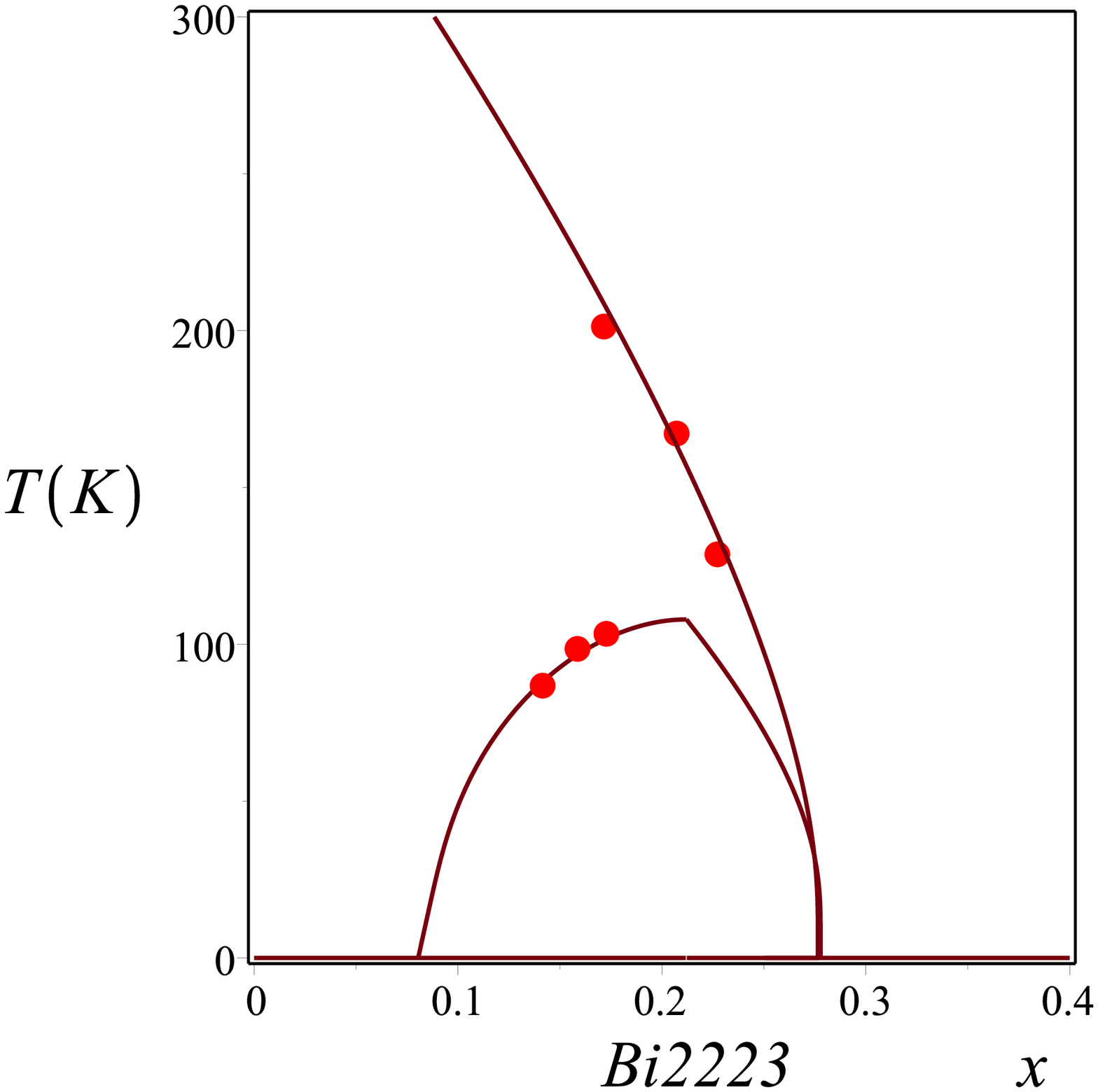}
	\caption{Solution of  Eqs. (\ref{ud}) and (\ref{od})  for the SC dome of Bi2223, together with the solution of (\ref{19a}) for the pseudogap temperature $T^*(x)$. Experimental data for $T_c(x)$ and for $T^*(x)$ both from\cite{7,8}.} \label{fige}
\end{figure}

\begin{figure}[!ht]
	\centering
	\includegraphics[scale=0.30]{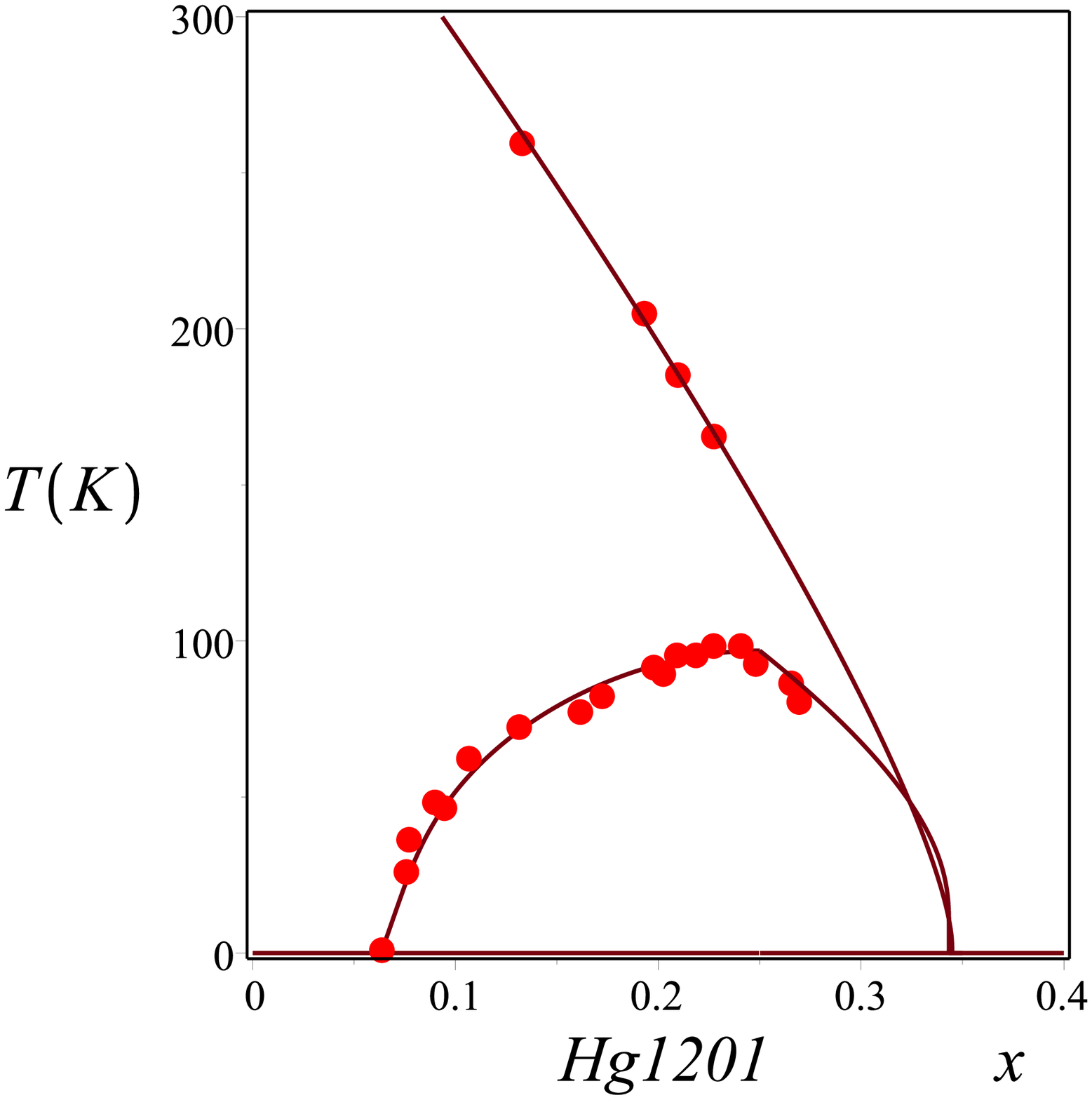}
	\caption{Solution of  Eqs. (\ref{ud}) and (\ref{od})  for the SC dome of Hg1201, together with the solution of (\ref{19a}) for the pseudogap temperature $T^*(x)$. Experimental data for $T_c(x)$ and for $T^*(x)$ both from\cite{7,8}.} \label{figc}
\end{figure}

\begin{figure}[!ht]
	\centering
	\includegraphics[scale=0.30]{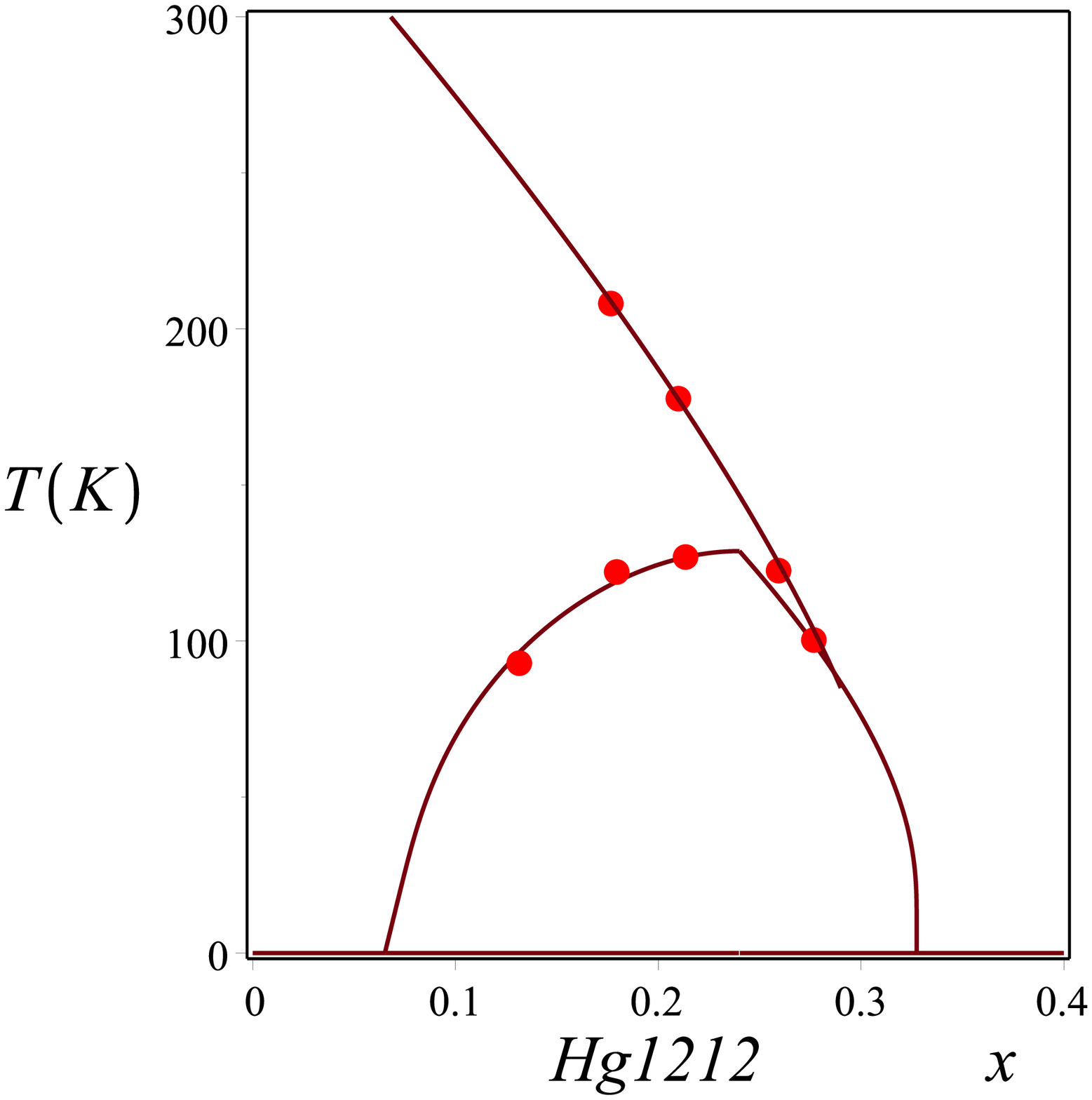}
	\caption{Solution of  Eqs. (\ref{ud}) and (\ref{od})  for the SC dome of Hg1212, together with the solution of (\ref{19a}) for the pseudogap temperature $T^*(x)$. Experimental data for $T_c(x)$ and for $T^*(x)$ both from\cite{7,8}.} \label{figf}
\end{figure}
\begin{figure}[!ht]
	\centering
	\includegraphics[scale=0.30]{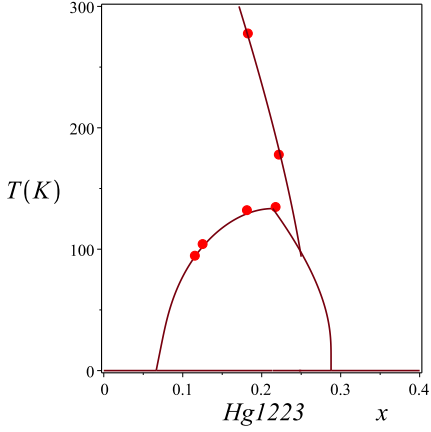}
	\caption{Solution of  Eqs. (\ref{ud}) and (\ref{od})  for the SC dome of Hg1223, together with the solution of (\ref{19a}) for the pseudogap temperature $T^*(x)$. Experimental data for $T_c(x)$ and for $T^*(x)$ both from\cite{7,8}.} \label{figg}
\end{figure}
\vfill\eject
{\bf 5.4) LSCO}\\
\bigskip

As we stated above, LSCO is apparently the only High-Tc cuprate, for which the curve $T_c(x)$ forms a dome, which is symmetric about $x_0$. Then, we must use now (\ref{udLSCO}) for determining the SC critical temperature
\begin{figure}[!ht]
	\centering
	\includegraphics[scale=0.30]{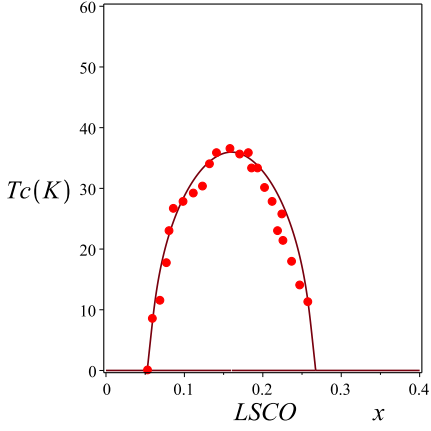}
	\caption{Solution of Eq. (\ref{udLSCO}) for the SC dome of LSCO. Experimental data from \cite{1,2,3,4}.} \label{fig1}
\end{figure}
\begin{figure}[!ht]
	\centering
	\includegraphics[scale=0.30]{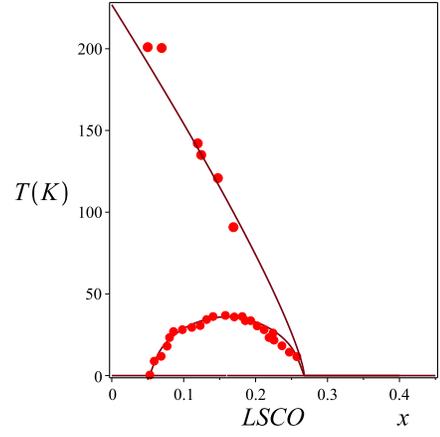}
	\caption{Solution of Eq. (\ref{udLSCO}) for the SC dome of LSCO, together with the solution of (\ref{19a}) for the pseudogap temperature $T^*(x)$. Experimental data for $T_c(x)$ from \cite{1,2,3,4} and for $T^*(x)$ from \cite{77}.} \label{figa}
\end{figure}

%The solution $T_c(x)$ of this implicit equation for the critical temperature of the SC transition, obtained with MAPLE, is depicted in Fig. \ref{fig1}.

\begin{figure}[!ht]
	\centering
	\includegraphics[scale=0.35]{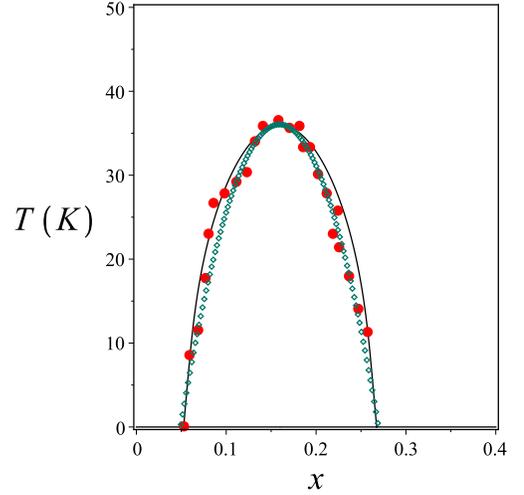}
	\caption{The empirical parabolic fit for the SC dome of LSCO, dotted line, superimposed with our solution for Eq. (\ref{19a}), solid line. Experimental data from \cite{1,2,3,4,7}.} 
	\label{fig2b}
\end{figure}

It is instructive to compare our result with the empirical curve, known since a long time \cite{honma,honma1,emp}, obtained by fitting the data for the LSCO dome, by the following parabola 
$$
T_c(x) =T_{max}\left[1- 82.616 (x_0-x)^2\right ].
$$
In Fig. \ref{fig2b} we superimpose it with our solution of
(\ref{udLSCO}).

\bigskip
%
%%%%%%%%%%%%%%%%%%%%%%%%%
\vfill\eject
{\bf 5.5) Universal Electronic Phase Diagram}\\
\bigskip

 The existence of a universality in the phase diagram of hole-doped cuprates has been reported in \cite{honma1}. This universality consists in the observation that, many hole-doped compounds have the same overall shaped phase diagram when expressed in terms of the variables $\tau_c\equiv T_c/T_{max}$ and $p=x/x_0$.
We can simply explain this result by using our expressions for $T_c(x)$ and $T_{max}$. Indeed, from

	\begin{equation}
		\tau_c(p)=
	\begin{cases}
		\frac{\ln2}{\ln2 + \frac{\zeta(1-p)}{\tau_{c}(p)} + \frac{1}{2}\left(e^{-\frac{2\zeta(1-p)}{\tau_{c}(p)}}-1\right)}, p<1\\
		\frac{\ln2 }{\ln\Big [1+ \exp\left[ \frac{-2\zeta(1-p)}{\tau_c(p)} \right]  \Big ]}, p>1
	\end{cases},
	\end{equation}
where the dimensionless factor $\zeta$, given by
	\begin{equation}
		\zeta=\frac{\gamma x_0}{T_{max}},
	\end{equation}
which is ultimately experimentally determined,
is the single parameter that governs the phase diagram.

	Using the parameters in Table \ref{t1}, we obtain the values of $\zeta$ for different compounds (see Table \ref{tab_zeta}).
	\begin{table}[!ht]
\begin{tabular}{|c|c|}
\hline\hline
           &$\zeta$\\ \hline \hline
Bi2201 & 1.16\\ %\hline
Bi2212  & 1.256\\% \hline
Bi2223   & 1.117\\ \hline
Hg1201 & 0.928\\% \hline
Hg1212 & 0.951\\ %\hline T_{max}=0.0108
Hg1223 & 1.005 \\ \hline
LSCO    & 1.032\\ \hline
\end{tabular}
\caption{$\zeta$ values for the families studied through this section.}
\label{tab_zeta}
\end{table}

\begin{figure}[!ht]
		\centering
		\includegraphics[width=\linewidth]{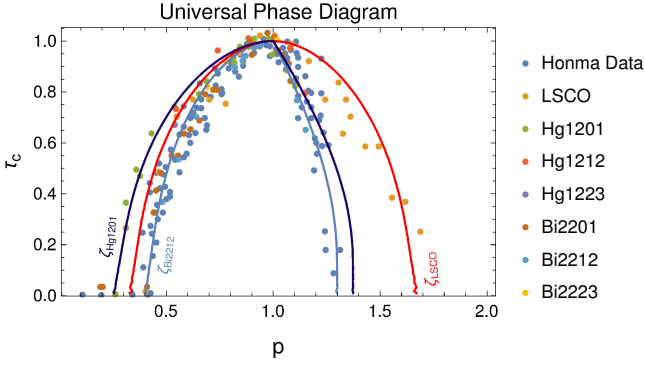}
		\caption{Universal Phase Diagram for all compounds studied together with the data on Ref. \cite{honma1}.}
		\label{fig_univ}
	\end{figure}

	In Fig. \ref{fig_univ} we show the phase diagrams for several compounds, including basically all the data of \cite{honma1}. These are essentially constrained between the values of $\zeta\in [0.951,1.256]$.

	The universality on the domes' shapes, when expressed in terms of $\tau(p)$ is a strong evidence that our Equations \ref{ud} and \ref{od} indeed correctly descibe the doping dependence of $T_c$ in hole-doped cuprates.\\
\bigskip

$$
\ \ \\ \ \ 
$$

{\bf 6) Increasing of $T_{max}$ with the Number of Layers  }\\
\bigskip

 It is an evident experimental fact that the optimal transition temperature becomes higher as one increases the number of   $CuO_2$ planes per primitive unit cell. Bi2201 and Hg1201, for instance, are single-layered materials, which have multi-layered relatives with a higher optimal temperature. 

The mercury family, for instance, consists of \cite{7,8,mer1,mer2}: Hg1201 (single-layered) ($T_{max} = 97\ K$), Hg1212 (double-layered) ($T_{max} = 125\ K$), Hg1223 (triple-layered) ($T_{max} = 134\ K$), Hg1234 (four-layered)($T_{max} = 127\ K$) and Hg1245 (five-layered)($T_{max} = 120\ K$).  It shows an increase of the optimal temperatures as the number of adjacent layers is increased from $N=1$ to $N=3$. Then for $N=4,5$, $T_{max} $ stabilizes at a temperature approximately corresponding to $N=2$ (see Table \ref{t3}) .

The same happens for the thallium family, for which \cite{honma,honma1,x},  Tl2201 (single-layered) ($T_{max} = 89\ K$), Tl2212 (double-layered) ($T_{max} = 119\ K$), Tl2223 (triple-layered) ($T_{max} = 128\ K$), Tl2234 (four-layered)($T_{max} = 119\ K$)(see Table \ref{t5}). 

For the bismuth family, accordingly, we have \cite{honma,honma1,bis1,bis2}
Bi2201 (single-layered) ($T_{max}$ =
34\  K), Bi2212 (double-layered) ($T_{max}$ = 92\  K), Bi2223
(triple-layered) ($T_{max}$ = 108\  K) (see Table \ref{t4}).

 From (\ref{14a}), we see that, for all members of a family, we may express the optimal temperature of a multi-layered cuprate with $N$ adjacent $CuO_2$ planes in terms of the corresponding temperature of the  single-layered one, as
\begin{eqnarray}
T_{max} (N)=\frac{\eta(Ng_S)}{\eta(g_S)} T_{max} (1).
\label{21}
\end{eqnarray}
 Observing that $\eta(N)$ is a monotonically increasing function of $N$, the obvious effect of increasing the number of adjacent planes is to increase $T_{max}$.
This follows directly from the enhancement of the coupling parameter, namely: $g  \rightarrow N g$. 
 
It is reasonable to admit that the coupling parameter $g_S$ is the same for all members of the same multi-layered family. In this case,  
     $\eta(Ng_S)$ can be expressed in terms of $\eta(g_S)$ using (\ref{eta3}).

We present a summary of the values of $\eta(Ng_S)$, as well as the predicted values of
$T_{max} (N)$, according to our model and the corresponding experimental values in Tables \ref{t3}, \ref{t5}  and \ref{t4}, respectively, for the bismuth, thalium and mercury families of cuprates.

\begin{table}[]

\begin{tabular}{|c|c|c|c|c|}
\hline\hline
           &  N           &  $ \eta(Ng_S)$ &   $T^{th}_{max}$ (K) & $T^{exp}_{max}$ (K)  \\ \hline\hline
Hg1201  &  1           &    0.61577    &          (96.8)                      &         96.8                       \\ \hline%\hline
Hg1212    &  2         &   0.80788       &           126.99           &           127                      \\ \hline
Hg1223     &   3       &    0.87192   &              137.07                   &         138                      \\ \hline
Hg1234     &   4       &    0.90394     &              142.10                &         127                      \\ \hline
Hg1245    &   5      &    0.92315   &                145.12                       &         120                      \\ \hline
\end{tabular}
\caption{The theoretical prediction of the optimal temperature as a function of the number of planes $N$ and the experimental values from \cite{honma,honma0,honma1,x} for the $Hg$ family}
\label{t4}
\end{table}

\begin{table}[]
\begin{tabular}{|c|c|c|c|c|}
\hline\hline
           &  N           &   $\eta(Ng_S)$  &   $T^{th}_{max}$ (K) & $T^{exp}_{max}$ (K)  \\ \hline\hline
Tl2201  &  1           &    0.59731         &          (89)                      &         89                       \\ \hline
Tl2212    &  2         &   0.79865        &            118.99                   &          119                      \\ \hline
Tl2223     &   3       &    0.86576     &              128.99                 &         128                      \\ \hline
Tl2234    &     4      &   0.89932     &               133.99                     &           119                      \\ \hline     
\end{tabular}
\caption{The theoretical prediction of the optimal temperature as a function of the number of planes $N$ and the experimental values from \cite{honma,honma0,honma1} for the $Tl$ family}
\label{t5}
\end{table}

\begin{table}[]
\begin{tabular}{|c|c|c|c|c|}
\hline\hline
           &  N           &   $\eta(Ng_S)$  &   $T^{th}_{max}$ (K) & $T^{exp}_{max}$ (K)  \\ \hline\hline
Bi2201  &  1           &    0.23077          &          (34.8)                      &         34.8                       \\ \hline
Bi2212    &  2         &   0.61538         &             92.79                   &          92.8                      \\ \hline
Bi2223     &   3       &    0.74358     &              112.13                       &         107.9                      \\ \hline
Bi2234    &     4      &   0.80769     &               121.80                     &           110                      \\ \hline     
\end{tabular}
\caption{The theoretical prediction of the optimal temperature as a function of the number of planes $N$ and the experimental values from \cite{honma,honma0,honma1} for the $Bi$ family}
\label{t3}
\end{table}

 The values  of $\eta(Ng_S)$ correspond to our theoretical expression (\ref{eta3}), whereas those in the column $T^{th}_{max}$ (K)
are obtained from (\ref{21}) and should be compared to the
 experimental values \cite{honma,honma1,bis1,bis2,mer1,mer2}, appearing on the last column.
\begin{figure}[!ht]
	\centering
	\includegraphics[scale=0.30]{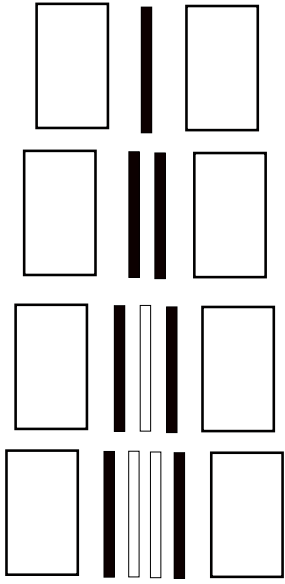}
	\caption{The $N$ $CuO_2$ planes (we show the cases where $N=1,...,4$) are squeezed between the charge reservoirs (white rectangles). For $N \geq 3$, there will be $N-2$ planes (white) without direct contact with these and therefore will be poorly doped.}
\label{figN}
\end{figure}

We see that
our theoretical values for the optimal temperature of the multi-layered members of the Bi and Hg families, are in good agree agreement with the experimental values for $N=2$. Then, for $N=3$, the agreement is within approximately $1\%$, whereas for $N>3$, there is no agreement.
The discrepancy, which starts to show at $N=4$ and increases for larger $N$'s can be ascribed to another effect that evidently must be taken into account as we increase the number of planes. This is the distance of such planes to the charge absorbing atoms doped into the system, which becomes progressively larger as the number of planes increases. Indeed, for $N=1$ we have the two ``charge reservoir'' regions adjacent to the unique $CuO_2$ plane. For $N=2$ still each of the two planes is adjacent to a charge reservoir. Then, for $N=3$ one of the planes is no longer adjacent to any charge reservoir, while for $N=4$ and $N=5$ the innermost planes are located far away from the charge reservoirs. It happens that while the outer planes are optimally doped the inner planes are poorly doped and, consequently, remain, to a large extent, underdoped \cite{ml}. The number of active $CuO_2$ planes, namely, the ones that are adjacent to a charge reservoir, in this case, is equivalent to the one we have for $N=2$, hence the temperature stabilizes at values similar to the ones we had for $N=2$.
\\

\bigskip

{\bf 7)  Effects of an Applied External Pressure}\\
\bigskip

{\bf 7.1)  Preliminary Considerations}\\
\bigskip

Under the effect of a change in pressure, given by $\Delta P = P-P_0$ a linear segment with original length $L_0$ would shrink to $L$, such that
\begin{eqnarray}
\frac{L- L_0 }{L_0}= - \kappa_L \Delta P,
\label{p1}
\end{eqnarray}
where $\kappa_L$ is the linear modulus of compressibility. For an infinitesimal change of pressure, $dP$, $L(P)$ would satisfy the linear differential equation, 
\begin{eqnarray}
\frac{1}{L}\frac{d L}{d P}= - \kappa_L ,
\label{p2}
\end{eqnarray}
which is solved by
\begin{eqnarray}
 L(P) = L_0 e^{ - \kappa_L   P }.
\label{p3}
\end{eqnarray}

 We would like to know how the SC critical temperatures are modified under the action of an external pressure. The crucial step for that comes from the connection we made between the coupling parameter $g_S$ of our effective model for the cuprates, and the magnetic exchange couplings of the model from which we started. Indeed, recent studies \cite{exch} have investigated how the magnetic coupling exchange integrals $J$ in the cuprates behave, under a change of pressure. From these, and from their connection with the $g$-couplings we can find how the $g_S$-parameter-dependent quantities behave as we change the external pressure.

It has been shown, in particular, that under a pressure variation $\Delta P$, the magnetic exchange coupling parameters behave as follows \cite{exch}:
\begin{eqnarray}
J(P)- J(P_0)= - \kappa_1\left (\frac{L- L_0 }{L_0}\right) .
\label{p4}
\end{eqnarray}
where $\kappa_1$ is a constant.

Using (\ref{p1}), therefore, we can write
\begin{eqnarray}
J(P)- J(P_0)=  \kappa_1 \kappa_L \Delta P ,
\label{p5}
\end{eqnarray}
from which we can define a modulus of compressibility for $J(P)$, namely,
\begin{eqnarray}
\frac{J(P)- J(P_0) }{J_0}=  \kappa_J \Delta P,
\label{p1a}
\end{eqnarray}
where $\kappa_{J}=\frac{\kappa_1 \kappa_L}{J_0}>0$. 

For an infinitesimal variation of pressure, this can be written, similarly to (\ref{p2}), as
\begin{eqnarray}
\frac{1}{J}\frac{dJ(P)}{dP}= \kappa_J.
\label{p6}
\end{eqnarray}
Solving this equation for $J(P)$, we obtain
\begin{eqnarray}
J(P)=J(0) e^{ \kappa_{J} P}.
\label{ap7}
\end{eqnarray}

The above expressions hold for AF couplings, when $J > 0$. The exponential dependence of $J$ on the pressure is intuitive as the exchange couplings involve the overlap between exponentially decaying wave-functions.

Assuming the muduli of compressibility, $\kappa_J$'s, for  the different magnetic couplings, $J_{AF}$, $J_K$,  are approximately the same and considering (\ref{01}), we come to the conclusion, that the SC coupling parameter $g_S$ grows exponentially with the pressure, with an effective modulus of compressibility, $\kappa_g$, which must be determined:
\begin{eqnarray}
g_S(P)=g_S e^{ \kappa_{g} P}
\label{p8}
\end{eqnarray}

Notice that $g_P=\frac{2t^2_p}{U_p}$ does not depend on the pressure. This happens because $U_p$ has twice as much overlap integrals as $t_p$ and therefore has a pressure dependence which goes as the square of  that of $t_p$. Hence we conclude that the pseudogap phenomena would not be influenced by the application of an external pressure.

In what follows we use our model and the above results for the pressure dependence of the coupling parameter to analyze the effect of pressure in the SC transition temperatures of cuprates. We first consider $T_{max}$ as it only involves the change of $\eta$ with pressure. Then we analyze how $T_c(x,P)$ changes as a function of pressure, for a fixed value of doping, as well as how the SC dome is modified for a fixed value of the pressure.\\
\bigskip
%\vfill\eject
$$
\ 
$$
{\bf 7.2) Variation of $T_{max}$  with the External Pressure}\\
\bigskip

From  (\ref{14a}), we see that the optimal SC transition temperature depends on pressure through the function $\eta(Ng_S(p))$. Then, inserting (\ref{p8}) into (\ref{eta1}),  we obtain the following curves for $T_{max}(P)$, respectively, for $Hg1212$ and $Hg1223$, after adjusting the parameter $\kappa_g$ to the single value $\kappa_{g}=\frac{1}{17}GPa^{-1}$ for both compounds.

\begin{figure}[!ht]
	\centering
	\includegraphics[scale=0.30]{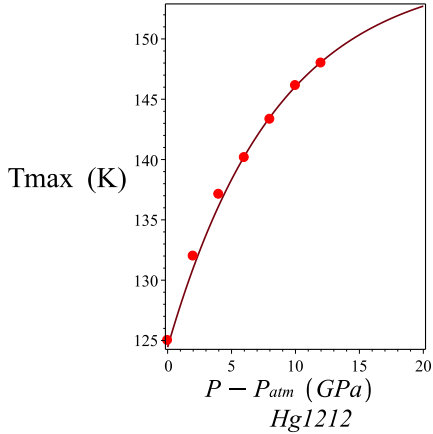}
	\caption{Optimal temperature of Hg1212 as a function of pressure, according to our theoretical prediction. Experimental data from \cite{p}.} \label{fig2p}
\end{figure}

\begin{figure}[!ht]
	\centering
	\includegraphics[scale=0.30]{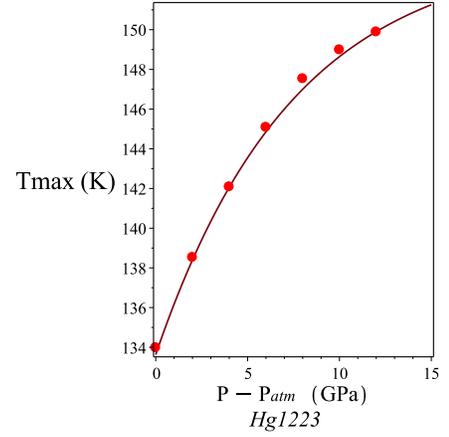}
	\caption{Optimal temperature of Hg1223 as a function of pressure, according to our theoretical prediction. Experimental data from \cite{p}.} \label{fig2pp}
\end{figure}

The fact that a single adjustment for  $\kappa_g$ works for both compounds indicates that the overlaps occurring in the exchange integrals do not change very much with the inclusion of more planes.

\begin{figure}[!ht]
	\centering
	\includegraphics[scale=0.60]{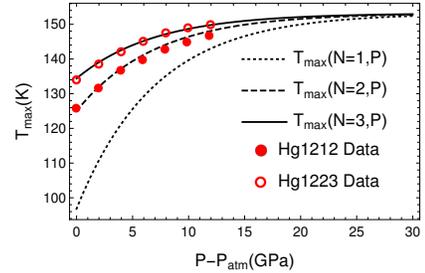}
	\caption{Optimal Temperature calculated from our model for the mercury family for one, two and three planes and comparison with data from \cite{p}.} \label{figTmaxall}
\end{figure}

In Fig. \ref{figTmaxall}, we compare the results for Hg1212 and Hg1223 and the prediction for Hg1201. We see that 
$T_{max}(P)$ saturates at a maximum value given by $\frac{\Lambda}{2\ln 2}$. This occurs because the optimal temperature depends on the coupling through the function $\eta(Ng_S)$ which itself saturates at $1$ as the coupling increases.
\\
\bigskip

{\bf 7.3) Variation of $T_{c}(x)$ with the External Pressure}\\
\bigskip

	In order to obtain the SC phase diagrams $T_c \times x$ for different values of the pressure, we take  $T_c(x)$ and adjust the value of the parameters $\gamma$ and $x_0$, for $P$ in the range $2-12 \ GPa$. The resulting values are displayed in Figs.\ref{x}, \ref{g}. The SC phase diagrams for $Hg1212$ and $Hg1223$, corresponding to $T_c(x,P)$ for different pressures  are shown below.

	\begin{figure}[!ht]
		\centering
		\includegraphics[scale=0.70]{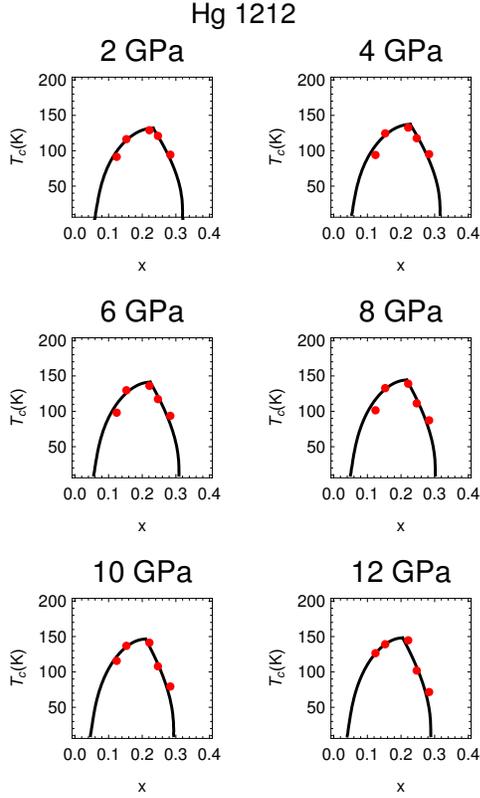}
		\caption{Phase diagram of Hg1212 as a function of pressure. Experimental data from \cite{p}. Solid line is our theoretical prediction.} \label{fig_diag_press_Hg1212}
	\end{figure}

	\begin{figure}[!ht]
		\centering
		\includegraphics[scale=0.70]{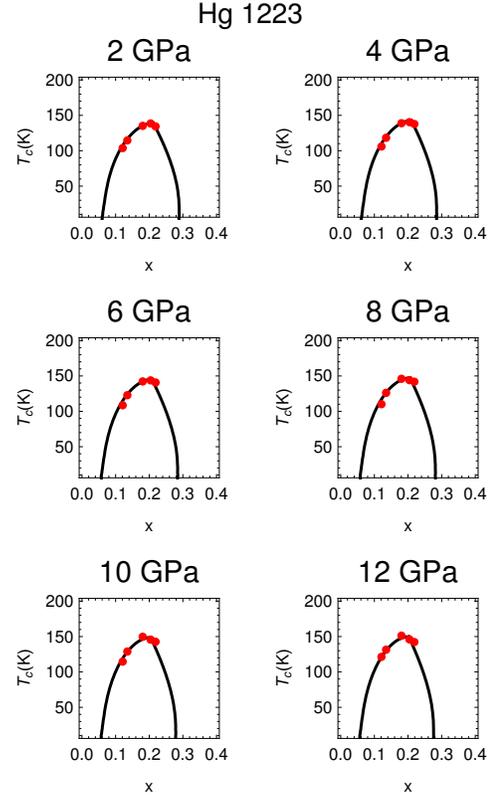}
		\caption{Phase diagram of Hg1223 as a function of pressure. Experimental data from \cite{p}. Solid line is our theoretical prediction.} \label{fig_diag_press_Hg1223}
	\end{figure}

The experimental data of the two previous figures were obtained from \cite{p}, using our equations (\ref{ud}) and (\ref{od}), instead of a parabola for obtaining the different doping values corresponding to each value of $T_c$ following the same procedure as in  \cite{p}.

	\begin{figure}[!ht]
		\centering
\includegraphics[scale=0.50]{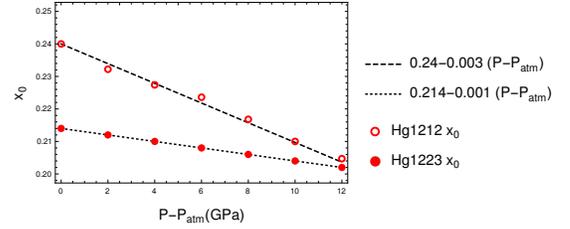}
				\caption{Dependence of $x_0$ on pressure for Hg1212 and Hg1223.} \label{x}
	\end{figure}

\begin{figure}[!ht]
		\centering
		\includegraphics[scale=0.50]{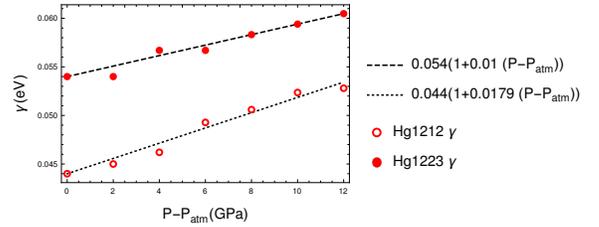}
		\caption{Dependence of $\gamma$ on pressure for Hg1212 and Hg1223.} \label{g}
	\end{figure}

We see a general trend in the figures above: $x_0$ decreases, $\gamma$ increases and the phase diagram becomes narrower for increasing values of pressure.

Observe that $\gamma$ and $x_0$ have an almost linear dependence on pressure, at least for the range of pressures considered. Using these values, we can obtain the pressure dependence of $T_c(x,P)$ for a fixed value of doping.

 For a fixed value of $x$ there is in general no monotonic increase of $T_c$ as a function of pressure as it happens with $T_{max}$. This occurs because $x-x_0(P)$ will change sign, depending on the value of $x$.

	To illustrate the different types behavior we select three situations at atmospheric pressure: one in the underdoped regime, $x<x_0(P=0)$ , one at the optimal doping, $x=x_0(P=0)$ and finally, one  in the overdoped regime $x>x_0(P=0)$.

\begin{figure}[!ht]
		\centering
		\includegraphics[scale=0.40]{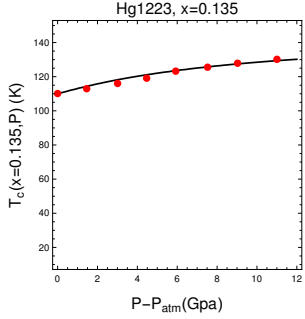}
		\caption{$T_c$ as a function of pressure for the $x=0.135$ compound. The data were taken from \cite{p}.} \label{fig_Tcx0135}
	\end{figure}

\begin{figure}[!ht]
	\centering
	\includegraphics[scale=0.30]{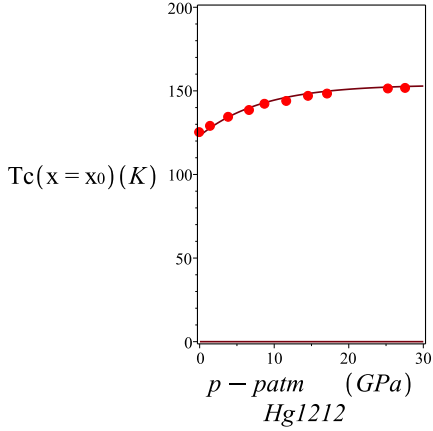}
	\caption{Temperature $T_c(x=x_0)$ of Hg1212 as a function of pressure. Experimental data from \cite{p2}.} \label{fig2ppp}
\end{figure}

\begin{figure}[!ht]
		\centering
		\includegraphics[scale=0.40]{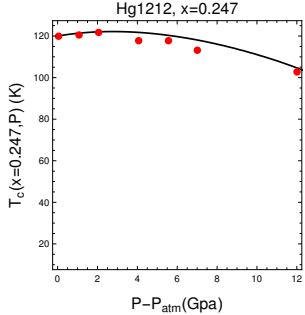}
		\caption{$T_c$ as a function of pressure for the $x=0.247$ compound. The data were taken from \cite{p}.} \label{fig_Tcx0247}
	\end{figure}

	In Fig. \ref{fig_Tcx0135} we present the results for an underdoped Hg1223, in Fig. \ref{fig2ppp}, for the optimal doped  Hg1212 and in  Fig. \ref{fig_Tcx0247} we plot $T_c$ for the slightly overdoped material Hg1212 x=0.247. 

Our theory for the pressure dependence of the temperature in High-Tc cuprates, which relies heavily on the connection of the SC coupling with the magnetic exchange couplings, has an excellent agreement with the experimental data. This adds more evidence for the correctness of our results.

Before concluding, we would like to mention recent experimental work that observes a re-entrance in Bi2212, exhibiting increasing Tc for a range of pressures much beyond the ones considered here, where a breakdown of the Fermi surface occurs \cite{break_press}. Despite the interest, this result is ouside the scope of our study.

{\bf 8) Concluding Remarks}\\
 We report here the obtainment of analytical expressions for $T_c(x)$, $T^*(x)$ and $T_c(x,P)$, for several High-Tc cuprate compounds, which show an excellent agreement with the experimental data. Our starting point is a Spin-Fermion-Hubbard (SFH) Model, describing the magnetic interactions among the localized spins of the copper ions, and the (Kondo) magnetic interactions between the localized and itinerant degrees of freedom (holes in oxygen p-orbitals). The model also describes the local Coulomb repulsion between the doped holes.

Our effective interaction among the doped holes contains two basic terms, defined on a bipartite oxygen latice: one of them is hole attractive while the other is hole repulsive.  These terms are derived, respectively, from the SFH model: a) by tracing out the localized degrees of freedom associated to the localized spins of the copper ions (see Appendix A) and; b) making a second order perturbation expansion in the hopping term in the Hubbard sector of the model (see Appendix C). Such interaction terms respectively favor the formation of Cooper pairs and excitons, and the condensation of each would lead to the SC and PG phases of the high-Tc cuprates. Each term respectively contains a coupling parameter $g_S$ or $g_P$, which, on one hand, can be expressed in terms of the parameters of the original Three Bands Hubbard Model and, on the other hand can be determined from the fit to the experimental data of the cuprates phase diagram. Remarkably, the values obtained by the two methods coincide.

By integrating over the fermions and minimizing the resulting effective action, we derive implicit equations, both for the SC transition temperature $T_c(x)$ and for the PG transition temperature $T^*(x)$, as a function of doping.   The solution of such equations is, then compared with the experimental data for different compounds, showing an excellent agreement, after the adjustment of a single parameter, which directly relates the chemical potential to the stoichiometric doping parameter $x$. The description of the chemical potential, directly made in terms of the stoichiometric doping parameter x, was a key factor for the success of our approach to the cuprates, since it has allowed to relate model calculated quantities, such as $T_c(x)$ and $T^*(x)$ with experimental data which are expressed exclusively in terms of x.

 The increase of $T_{max}$ with the number of adjacent planes in multi-layered cuprates can be understood as a consequence of the enhancement of the SC coupling $g_S\rightarrow Ng_S$ produced by the presence of these planes. As $N$ increases, however, the inner planes progressively recede from the charge reservoirs, an effect that counteracts the enhancement of the coupling parameter, thus leading to a stabilization (or even decrease) of $T_{max}$ as we increase $N$. 

Based on our results one can devise a way to increase $T_c$ in cuprates: this would be achieved by effectively doping the innermost planes in multilayered cuprates. For that purpose, one should design materials with a unit cell containing as many layers as possible but with charge reservoirs intercalating no more than two layers. This would neutralize the above effect, thereby increasing $T_c$.

We finally employ our results in order to analyze the effects of an applied pressure on $T_c(x)$, again, obtaining excellent agreement with experiments for $T_{max}(P)$ and for $T_{c}(x,P)$, both for fixed $x$ and fixed $P$. Based on our model we are also able to predict that the PG transition temperature will not be affected by the application of an external pressure.

Our results open new avenues for the investigation of the physical properties of high-Tc cuprates, with outstanding possibilities. Among these, employing our model for: the description of resistivity above $T_c$; the description of charge ordering phases;  the determination of the specific heat; the investigation of the detailed nature of the strange-metal phase at $T>T^*$; the study of the Fermi liquid phase in the far overdoped region.

 The study reported here consists in a concrete step forward in the attempt to understand high-Tc superconductivity. \\
\bigskip
\vfill\eject
{\bf Acknowledgments}

We thank A. V. Balatsky and C. Morais Smith for stimulating conversations. We would also like to thank E. Fradkin, S. Kivelson, J. Tranquada and P. A. Marchetti, for very useful comments.
E. C. Marino was supported in part by CNPq and by FAPERJ. R. Arouca acknowledges CNPq for support. V. S. Alves acknowledges CNPq for support. Reginaldo de Oliveira Jr acknowledges CAPES and FAPERJ for support. \\
\bigskip
\\
\bigskip

{\it Corresponding author: ECM (marino@if.ufrj.br)}

\appendix
\section{Appendix A}
% \def\thechapter {\Alph{chapter}}
% \def\thesection {\thechapter.\arabic{section}}
% \def\thetable   {\thechapter.\@arabic\c@table}
% \def\thefigure  {\thechapter.\@arabic\c@figure}
%\setcounter[equation]{0}
% \def\theequation{\thechapter.\arabic{equation}}

%\bigskip
Let us perform here the trace over the localized copper spin magnetic moments in the full partition function, which is given by (\ref{a}). This trace only runs over the localized degrees of freedom, $\textbf{S}_I$  and can be expressed as
 \begin{eqnarray}
 {\rm Tr}_{\textbf{S}_I} e^{- \beta\Big[ H_{AF}[\textbf{S}_I]+ H_K[\textbf{S}_I,\psi]\Big ]}, 
\label{976}
\end{eqnarray}
where $H_{AF}$ and $H_{K}$ are given by (\ref{0a}).
By tracing only over the localized spin degrees of freedom $\textbf{S}_I$, we are able to obtain a contribution for effective interaction Hamiltonian of the doped holes, namely $H_{1}[\psi]$, which is defined from the relation
\begin{eqnarray}
e^{- \beta H_{1}[\psi]}\propto {\rm Tr}_{\textbf{S}_I} e^{- \beta\Big[ H_{AF}[\textbf{S}_I]+H_{K}[\textbf{S}_I,\psi]\Big]}, 
\label{976a}
\end{eqnarray}

In order to evaluate the trace above, we shall use the coherent spin states $|\textbf{N}\rangle$, (see, for instance (\cite{ecm2}) ) defined by the property
\begin{eqnarray}
\langle\textbf{N}|\textbf{S}|\textbf{N}\rangle= s \textbf{N}\ \ ;\ \ \int_{S^2}\frac{d\Omega}{4\pi}|\textbf{N}\rangle \langle\textbf{N}|= 1
\label{977}
\end{eqnarray}
where $s=1/2$ is the spin quantum number, $\textbf{N}$ is a unit classical vector and the integration is made over the whole solid angle.

In terms of this, we can express the partial trace over $\textbf{S}_I$  in (\ref{976a}) as a functional integral over $\textbf{N}$ (see, for instance \cite{ecm2}):

\begin{eqnarray}
& & {\rm Tr}_{\textbf{S}_I} e^{- \beta\Big[ H_{AF}[\textbf{S}_I]+H_{K}[\textbf{S}_I,\psi]\Big]} =
\nonumber \\
& & \int D \textbf{N}  \exp\left\{ \int_0^\beta d\tau \left [\langle\textbf{N}(\tau)|\frac{d}{d\tau}|\textbf{N}(\tau)\rangle
- H[s\textbf{N}] \right ]\right \}.\nonumber
\\
\label{976b}
\end{eqnarray}
Here
\begin{eqnarray}
 H[s\textbf{N}]= H_{AF}[s\textbf{N}]+H_{K}[s\textbf{N},\psi].
\label{976bb}
\end{eqnarray}

We now separate $\textbf{N}$ in antiferromagnetic and ferromagnetic components, denoted, respectively, by
 $\textbf{n}$ and $\textbf{L}$, such that $|\textbf{n}|^2=1$ and 
$\textbf{L}\cdot\textbf{n}=0$. We write, in site $I$, in terms of the lattice parameter $a$,
\begin{eqnarray}
\textbf{N}_I= (-1)^I \textbf{n}_I+a^2\textbf{L}_I+O(a^4)
\label{988}
\end{eqnarray}
in such a way that  $|\textbf{N}|^2=|\textbf{n}|^2=1$.

We can express the trace in (\ref{976a}), in the continuum limit, as a double functional integral on  $\textbf{n}$ and $\textbf{L}$ \cite{ecm2}:
\begin{eqnarray}
& &\hspace{-5mm}e^{- \beta  H_{int}[\psi]}= \int D\textbf{n}D\textbf{L}\delta(|\textbf{n}|^2-1)\times
\nonumber \\
&\ &\hspace{-5mm} \exp\left\{\frac{1}{2}\int d^2r\int_0^\beta d\tau\left[ J_{AF}s^2 \nabla_i \textbf{n}\cdot \nabla_i \textbf{n} \right. \right .
\nonumber \\
&\ &\hspace{-5mm}\left .\left .+ 4J_{AF}s^2a^2  |\textbf{L}|^2\right] + \textbf{L}\cdot\left[ J_K \mathcal{S}-is\textbf{n}\times\frac{\partial \textbf{n} }{\partial \tau} \right] \right\}\label{99110}
\end{eqnarray}
where 
$$
\mathcal{S}=\mathcal{S}_A +\mathcal {S}_B
$$
is given by (\ref{0b}).

We now integrate out the ferromagnetic fluctuations by performing the quadratic functional integral on $\textbf{L}$. This will produce the square of the last term between brackets, which contains three terms: the 2nd term squared, which provides a kinetic term for $\textbf{n}$ \cite{ecm2}, the crossed term, which vanishes \cite{ecm1} and the 1st term squared that yields a $\psi$-dependent interaction term. This consists basically of an effective AF magnetic interaction among the itinerant  doped holes.

\begin{eqnarray}
& &e^{- \int_0^\beta d\tau H_{int}[\psi]}=\int D\textbf{n}\delta(|\textbf{n}|^2-1) \times
\nonumber \\
& &\exp\left\{\int d^2r \int_0^\beta d\tau\frac{\rho_s}{2}\left[  \nabla_i \textbf{n}\cdot \nabla_i \textbf{n} +\frac{1}{c^2} \partial_\tau \textbf{n}\cdot \partial_\tau \textbf{n}
\right]\right .
\nonumber \\
& &+\frac{J^2_K}{8J_{AF}a^2}\Big[\mathcal{S}\cdot \mathcal{S}    \Big ]
\left . \right . \Big\}
\label{99113}
\end{eqnarray}
where $\rho_s = \frac{J_{AF}}{4}$ is the spin stiffness and $c=J_{AF}a$ is the spin-waves velocity.

Using the fact that  the continuum limit involves the 
$a^2 \sum_k \leftrightarrow \int d^2r $, we conclude that
\begin{eqnarray}
& &e^{- \int_0^\beta d\tau H_{1}[\psi]}=
 {\rm Tr}_{\textbf{S}_I} e^{- \beta\Big[ H_{AF}[\textbf{S}_I]+H_{K}[\textbf{S}_I,\psi]\Big]} =
\nonumber \\
& & Z_{NL\sigma M}\exp\left\{\int_0^\beta d\tau\sum_{\textbf{R},\textbf{R}+\textbf{d}} \left[ \frac{J^2_K}{8J_{AF}} [\mathcal{S}_A +\mathcal{S}_B]^2\right ]\right\},
\nonumber \\
\label{976b}
\end{eqnarray}
where $Z_{NL\sigma M}$ is the partition function of the Nonlinear Sigma Model ( see, for instance \cite{ecm2}).

From the last term in (\ref{976b}) we see that, indeed,
$$
H_{1}[\psi]=-\frac{J^2_K}{8J_{AF}}\sum_{\textbf{R},\textbf{R}+\textbf{d}}  [\mathcal{S}_A +\mathcal{S}_B]^2.
 $$
as we find in (\ref{22a}).\\
\bigskip
\appendix
\section{ Appendix B}
%\setcounter
%\bigskip

In this Appendix, we demonstrate how to obtain $H_{SC}[\psi]$, out of (\ref{22a}).

Using the Pauli matrices, we can express the three components of the holes' spin as
\begin{eqnarray}
S^X=\frac{1}{2}\sum_{C=A,B}\Big[ \psi^\dagger_{C\uparrow} \psi_{C\downarrow}+ \psi^\dagger_{C\downarrow} 
\psi_{C\uparrow}\Big ]
\label{b1}
\end{eqnarray}

\begin{eqnarray}
S^Y=\frac{1}{2}\sum_{C=A,B}i\Big[ \psi^\dagger_{C\uparrow} \psi_{C\downarrow}- \psi^\dagger_{C\downarrow} 
\psi_{C\uparrow}\Big ]
\label{b2}
\end{eqnarray}

\begin{eqnarray}
S^Z=\frac{1}{2}\sum_{C=A,B}\Big[ \psi^\dagger_{C\uparrow} \psi_{C\uparrow}- \psi^\dagger_{C\downarrow} 
\psi_{C\downarrow}\Big ]
\label{b3}
\end{eqnarray}

Inserting these expressions in (\ref{22a}), 
and defining 
\begin{eqnarray}
& &n^C_{++}= \psi^\dagger_{C\uparrow} \psi_{C\uparrow} \ \ ;\ \ n^C_{--}=\psi^\dagger_{C\downarrow} \psi_{C\downarrow}
\nonumber \\
& &n^C_{+-}= \psi^\dagger_{C\uparrow} \psi_{C\downarrow} \ \ ;\ \ n^C_{-+}=\psi^\dagger_{C\downarrow} \psi_{C\uparrow},
\label{b4}
\end{eqnarray}
for $C=A,B$,
we can write $H_{int}$ as
\begin{eqnarray}
& \hspace{-3mm}H_{int}=&-\frac{J^2_K}{8J_{AF}}\sum_{\textbf{R},\textbf{R}+\textbf{d}}\frac{1}{4}\left \{ \Big[n^A_{+-}+n^A_{-+}  +n^B_{+-}+n^B_{-+} \Big]^2  \right .
\nonumber \\
& & \hspace{-3mm}-\Big[n^A_{+-}-n^A_{-+}  +n^B_{+-}-n^B_{-+} \Big]^2
\nonumber \\
& & \hspace{-3mm} \left .+\Big[n^A_{++}-n^A_{--}  +n^B_{++}-n^B_{--} \Big]^2 \right \}
\label{b5}
\end{eqnarray}
Then, considering that we can rewrite (\ref{4a}) as
\begin{eqnarray}
& & \Sigma =n^A_{+-}n^B_{-+}+n^A_{-+}n^B_{+-}
\nonumber \\
& & \Pi=n^A_{+-}n^B_{+-}+n^A_{-+}n^B_{-+}
\nonumber \\
 & & \Xi_1 = n^A_{++}n^B_{--}+n^A_{--}n^B_{++}
\nonumber \\
 & & \Xi_2 = n^A_{++}n^B_{++}+n^A_{--}n^B_{--}
\label{b6}
\end{eqnarray}
and using (\ref{b5}), we establish (\ref{0}), up to a constant.
 $$
\\ 
$$
\bigskip
\appendix
\section{ Appendix C}
%\setcounter
%\\
%\bigskip

In this Appendix, we perform a perturbation expansion in $t_{p}/U_p$, in $H_0+H_U$, given by
\begin{eqnarray}
& &H_0=
%\nonumber \\
-t_p \sum_{\textbf{R},\textbf{d}_i}\Big [\psi_{A\sigma}^\dagger(\textbf{R}) \psi_{B\sigma}(\textbf{R}+\textbf{d}_i)  
 \nonumber \\
& &+\psi_{B\sigma}^\dagger(\textbf{R}+
\textbf{d}_i) \psi_{A\sigma}(\textbf{R}) + HC \Big ]
\nonumber \\
&\ &H_U= U_p \sum_{\textbf{R}} n^A_\uparrow n^A_\downarrow + U_p \sum_{\textbf{R}+\textbf{d}} n^B_\uparrow n^B_\downarrow 
\nonumber \\
\label{Ut}
\end{eqnarray}
by taking $H_0$ as a perturbation and $H_U$ as the unperturbed Hamiltonian.

 We are going to determine the ground state energy $E_0$, up to second order, as the series:
\begin{eqnarray}
E_0 = E_0^{(0)} + E_0^{(1)} + E_0^{(2)} ,
\label{Ut1}
\end{eqnarray}
where $E_0^{(0)}$ is the unperturbed ground state energy and $ E_0^{(1)}$ are, respectively, the first and second order corrections. We will then find an effective hamiltonian $H_{2}$ such that such that $E_0=\langle H_{2}\rangle$, namely, the first order correction to $H_{2}$ is given by $ (\ref{Ut1})$.

\begin{figure}[!ht]
	\centering
	\includegraphics[scale=0.25]{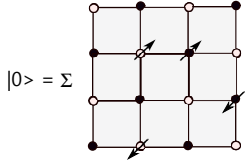}
	\caption{Unperturbed ground state.} \label{ef}
\end{figure}

The unperturbed ground state $|0\rangle$ (ground state of  $H_U$), depicted in Fig. \ref{ef}, satisfies
\begin{eqnarray}
H_0 |0\rangle = E_0^{(0)}|0\rangle = 0 \ \ \  \; \ \ \   \langle 0| H_0 |0\rangle =0
\label{Ut2}
\end{eqnarray}
thus implying that $E_0^{(0)} = E_0^{(1)}=0 $.

\begin{figure}[!ht]
	\centering
	\includegraphics[scale=0.25]{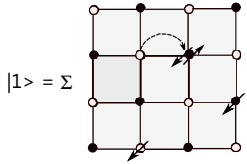}
	\caption{Unperturbed excited state.} \label{ee}
\end{figure}

The second order correction is given by
\begin{eqnarray}
 E_0^{(2)}= \sum_{n \neq 0}\frac{\langle 0| H_0|n\rangle\langle n|H_0 |0\rangle}{E_0^{(0)}-E_n^{(0)}} ,
\label{Ut3}
\end{eqnarray}
From Fig. \ref{ee}, we see that $|1\rangle$, such that
$$
H_0 |1\rangle = U_p |1\rangle 
$$
 is the only excited state contributing to (\ref{Ut3}), hence
\begin{eqnarray}
 E_0^{(2)}=-\frac{1}{U_p}\langle 0| H_0 H_0 |0\rangle ,
\label{Ut4}
\end{eqnarray}
Inserting the expression of $H_0$, taken from (\ref{Ut1}), we see that only the two crossed terms contribute
and 
\begin{eqnarray}
 E_0 = E_0^{(2)}=\langle 0| H_{2}[\psi] |0\rangle ,
\label{Ut5}
\end{eqnarray}
where $ H_{2}[\psi]$ is given by
\begin{eqnarray}
H_{2}[\psi]=- \frac{2t^2_p}{U_p}\Big[\psi_{A\sigma}^\dagger(\textbf{R}) \psi_{B\sigma}(\textbf{R}+\textbf{d}_i) \Big]
\Big[\psi_{B\sigma}^\dagger(\textbf{R}+\textbf{d}_i) \psi_{A\sigma}(\textbf{R}) \Big]
\nonumber \\.
\label{Ut6}
\end{eqnarray}
Equivalently, using (\ref{b6}), we can write
\begin{eqnarray}
H_{2}[\psi]= -\frac{2t^2_p}{U_p}\Big[\Pi-\Xi_2 \Big]
\label{Ut7}
\end{eqnarray}

We therefore establish
 (\ref{4a}), (\ref{4az}) and  (\ref{4b}).

\end{document}